\documentclass[12pt]{article}

\usepackage{amsmath}
\usepackage{graphicx,psfrag,epsf}
\usepackage{enumerate}
\usepackage{natbib}
\usepackage{url} 
\usepackage[ruled]{algorithm2e}
\usepackage{subcaption}
\usepackage{amssymb,amsthm}
\usepackage{hyperref}

\newcommand{\blind}{0}

\newcommand{\R}{\mathbb{R}}
\newcommand{\myc}{c}
\newcommand\abs[1]{\left\lvert#1\right\rvert}
\newcommand{\E}{\mathbb{E}}
\newcommand{\var}{\operatorname{var}}
\newcommand{\cov}{\operatorname{cov}}
\newcommand{\Xq}{X^q}
\newcommand{\Xminusq}{X^{-q}}
\newcommand{\pq}{p_q}
\newcommand{\Rq}{R_q}
\newcommand{\pqW}{\widehat{p_q}}
\newcommand{\RqW}{\widehat{R_q}}
\newcommand{\pqG}{\widehat{p_q}^\text{G}}
\newcommand{\pG}[1][q]{\widehat{p}_{#1}^\text{G}}
\newcommand{\RqMC}{\widehat{R_q}^\text{MC}}
\newcommand{\pGMC}{\widehat{p}^\text{GMC}}
\newcommand{\RqANMC}{\widehat{R_q}^\text{anMC}}
\newcommand{\pGANMC}{\widehat{p}^\text{GanMC}}
\newcommand{\RprogLang}{R}
\newcommand{\muGP}{\mathfrak{m}}

\newtheorem{proposition}{Proposition}
\newtheorem{corollary}{Corollary}

\date{}

\begin{document}

\def\spacingset#1{\renewcommand{\baselinestretch}%
{#1}\small\normalsize} \spacingset{1}


\if0\blind
{
  \title{\bf Estimating orthant probabilities of high dimensional Gaussian vectors with an application to set estimation}
  \author{Dario Azzimonti\thanks{
    The first author gratefully acknowledges the support of the \textit{ Swiss National Science Foundation, grant number 146354} and of the \textit{Hasler Foundation, grant number 16065.}}\hspace{.2cm}\\
UQOD group, Idiap Research Institute and \\
    IMSV, Department of Mathematics and Statistics, University of Bern \\
    and \\
    David Ginsbourger \\
    UQOD group, Idiap Research Institute and \\
    IMSV, Department of Mathematics and Statistics, University of Bern.}
  \maketitle
} \fi

\if1\blind
{
  \bigskip
  \bigskip
  \bigskip
  \begin{center}
    {\LARGE\bf Title}
\end{center}
  \medskip
} \fi

\bigskip
\begin{abstract}
The computation of Gaussian orthant probabilities has been extensively studied for low-dimensional vectors. Here, we focus on the high-dimensional case and we present a two-step procedure relying on both deterministic and stochastic techniques. The proposed estimator relies indeed on splitting the probability into a low-dimensional term and a remainder. While the low-dimensional probability can be estimated by fast and accurate quadrature, the remainder requires Monte Carlo sampling. We further refine the estimation by using a novel asymmetric nested Monte Carlo (anMC) algorithm for the remainder and we highlight cases where this approximation brings substantial efficiency gains. The proposed methods are compared against state-of-the-art techniques in a numerical study, which also calls attention to the advantages and drawbacks of the procedure. Finally, the proposed method is applied to derive conservative estimates of excursion sets of expensive to evaluate deterministic functions under a Gaussian random field prior, without requiring a Markov assumption. Supplementary material for this article is available online.
\end{abstract}

\noindent%
{\it Keywords:}  Conservative set estimation; Gaussian probabilities; Gaussian random fields; Monte Carlo.
\vfill
\hfill {\tiny This is an Accepted Manuscript of an article published by Taylor \& Francis Group in Journal of Computational and Graphical Statistics  on 03/08/2017, available online: \url{https://doi.org/10.1080/10618600.2017.1360781}}

\newpage
\spacingset{1.45} 
\section{Introduction}
\label{sec:Intro}

Assume that $X = (X_1, \dots, X_d)$ is a random vector with Gaussian distribution $N_d(\mu,\Sigma)$.
We are interested in estimating, for any fixed $t \in \R$, the following probability
\begin{equation}
\pi(t)= P(X \leq (t, \dots, t)).
\label{eq:probMax}
\end{equation}
The general problem of evaluating $\pi(t)$, which, for a full rank matrix $\Sigma$, is the integral of the multivariate normal density $\phi(\cdot;\mu,\Sigma)$ over the one-sided $d$-dimensional rectangle $(-\infty,t]^d$, has been extensively studied in moderate dimensions with many different methods. 
In low dimensions tables are available (see, e.g.,~\citet{owen1956tables} for $d=2$). 
Furthermore, when the dimension is smaller than $20$, there exist methods (see, e.g.,~\citet{abrahamson1964orthant},~\citet{moran1984monte},~\citet{miwa2003evaluation} and \citet{Craig2008orthant}) exploiting the specific orthant structure of the probability in~\eqref{eq:probMax}. Currently, however, most of the literature uses numerical integration techniques to approximate the quantity. In moderate dimensions fast reliable methods are established to approximate $\pi(t)$ (see, e.g.~\citet{cox1991simple}) and more recently the methods introduced in~\citet{schervish1984algorithm,genz1992numerical} and~\citet{hajivassiliou1996simulation} (see also~\citet{genz2002comparison},~\citet{ridgway2014computation} and the book~\citet{Genz.Bretz2009} for a broader overview) provide state-of-the-art algorithms when $d<100$. The method introduced by~\citet{genz1992numerical} has been recently revised in~\citet{botev2016normal} where a more efficient tilted estimator is proposed. Those techniques rely on fast quasi Monte Carlo (qMC) methods and are very accurate for moderate dimensions. Here we focus on problems where $d$ is larger than $1000$ and $\pi(t)$ is not a rare event probability. Such estimation problems occur, for example, if $\pi(t)$ comes from a discretization of a Gaussian random field and $t$ is a fixed finite threshold.  In such cases, existing techniques are not computationally efficient or become intractable. Commonly used alternative methods are standard Monte Carlo (MC) techniques (see~\citet{tong2012multivariate}, Chapter~8 for an extensive review), for which getting accurate estimates can be computationally prohibitive.

We propose here a two step method that exploits the power of qMC quadratures and the flexibility of stochastic simulation for the specific problem of estimating $\pi(t)$. 
We rely on the following equivalent formulation.
\begin{equation*}
\pi(t) = 1 - P(\max X >t),
\end{equation*} 
where $\max X$ denotes $\max_{i=1, \dots, d}X_i$. In the following we fix $t$ and denote $p=P(\max X >t)$.

The central idea is using a moderate dimensional subvector of $X$ to approximate $p$ and then correcting bias by MC. Let us fix $q \ll d$ and define the active dimensions as $E_q =\{{i_1}, \dots, {i_q}\} \subset \{1, \dots, d\}$. 
Let us further denote with $\Xq$ the $q$ dimensional vector $\Xq= (X_{i_1}, \dots, X_{i_q})$ and with $\Xminusq$ the $(d-q)$ dimensional vector $\Xminusq= (X_j)_{j \in E\setminus E_q}$. 
Then, 
\begin{align}
\label{eq:pMaxDecom}
p = P(\max X > t) &= \pq + (1-\pq)\Rq, 
\\ \nonumber
\pq &= P(\max\Xq > t), \\ \nonumber
\Rq &= P(\max\Xminusq > t \mid \max\Xq \leq t).
\end{align}
The quantity $\pq$ is always smaller or equal to $p$ as 
$E_q \subset \{1, \dots, d\}$.
Selecting a non-degenerate vector $\Xq$, we propose to estimate $\pq$ with the QRSVN algorithm \citep{Genz.etal2012} which is efficient as we choose a number of active dimensions $q$ much smaller than $d$.
In~\citet{Chevalier2013}, Chapter~6, the similar problem of approximating the non-exceedance probability of the maximum of a Gaussian random field (GRF) $\xi$ based on a few well-selected points is presented. Each component of $X$ stands for the value of $\xi$ at one point of a given discretization of the field's domain. Active dimensions 
(i.e. the well-selected points) 
were chosen by numerically maximizing $\pq$, and the remainder was not accounted for. 
Our proposed method, instead, does not require a full optimization of the active dimensions as we exploit the decomposition in~\eqref{eq:pMaxDecom} to correct the error introduced by $\pq$. For this task, we propose two techniques to estimate the reminder $\Rq$: a standard MC technique and a novel asymmetric nested Monte Carlo (anMC) algorithm. The anMC technique draws samples by taking into account the computational cost, resulting in a more efficient estimator. 

The anMC method presented is quite general, however its overall performance is depends on the techniques chosen to estimate $\pq$ and $\Rq$. The choices described in this paper are implemented as default in the $\RprogLang$ programming language~\citep{Rcore2017} in the package \verb|anMC|, however numerical experiments presented in Appendix~\ref{sec:NumSmallProba} and in supplementary material show that, for some specific problems, alternative choices might be better suited.


In the remainder of the paper, we propose an unbiased estimator for $p$ and we compute its variance in Section~\ref{sec:MainTheory}. 
In Section~\ref{sec:ANMC} we introduce the anMC algorithm in the more general setting of estimating expectations depending on two vectors with different simulation costs. It is then explicitly applied to efficiently estimate $\Rq$. In Section~\ref{sec:NumStudies} the results of two numerical studies are reported. The first one studies the efficiency of the anMC algorithm compared to standard MC. The second one is a benchmark study where the efficiency of the proposed methods is compared with a selection of state-of-the-art techniques. This study is extended to the case of small and very high probabilities in Appendix~\ref{sec:NumSmallProba}. In Section~\ref{sec:ConservativeEsts}, we present an implementation of this method to compute conservative estimates of excursion sets for expensive to evaluate functions under non-necessarily Markovian Gaussian random field priors. More details on the choice of active dimensions are presented in Appendix~\ref{sec:ActiveDims}. All proofs are in Appendix~\ref{sec:proofs}. Computer code for partially replicating the experiments presented here is attached in supplementary material, where we also report the results of an additional numerical experiment and a study on the computational times for the application of Section~\ref{sec:ConservativeEsts}. The figures summarizing the benchmark results were produced with the package~\verb|ggplot2|~\citep{Wickham_ggplot2}.

\section{The estimator properties}
\label{sec:MainTheory}

\subsection{An unbiased estimator for $p$}
Equation~\eqref{eq:pMaxDecom} gives us a decomposition that can be exploited to obtain an unbiased estimator for $p$. In the following proposition we define the estimator and we compute its variance.
%

\begin{proposition}
	Consider $\pqW$ and $\RqW$, independent unbiased estimators of $\pq$ and $\Rq$ respectively, then $ \widehat{p} = \pqW + (1-\pqW)\RqW$ is an unbiased estimator for $p$. Moreover its variance is
	\begin{equation}
	\var(\widehat{p}) = (1 - \Rq)^2\var(\pqW) + (1 - \pq)^2\var(\RqW) + \var(\pqW)\var(\RqW).
	\label{eq:VarHatp}
	\end{equation}
	\label{pro:VarHatp}
\end{proposition}

This property is independent from the choice of estimators $\pqW$ and $\RqW$. In what follows we consider different efficient computational strategies for $\pqW$ and $\RqW$.



\subsection{Quasi Monte Carlo estimator for $\pq$}
\label{subsec:pq}

The quantity $\pq$ can also be computed as
\begin{equation*} 
\pq = 1 - P\left(\Xq \leq t_q\right),
\end{equation*}
where $t_q$ denotes the $q$ dimensional vector $(t, \dots, t)$. 
The approximation of $\pq$ thus requires only an evaluation of the c.d.f. of $\Xq$. 
We denote with $\pqW$ a generic estimator for $\pq$ and, since we assume that $q \ll d$, we propose to estimate $\pq$ with the estimator $\pqG$ that uses the randomized quasi Monte Carlo integration method $\mathrm{QRSVN}$ introduced in~\citet{genz1992numerical},~\citet{hajivassiliou1996simulation} and refined in~\citet{Genz.Bretz2009}. 
In particular we consider here the implementation of QRSVN with the variable reordering described in~\citet[Section~4.1.3]{Genz.Bretz2009}. 
The estimate's error is approximated with the variance of the randomized integration. 
The quantity $\pqG$ obtained with this procedure is an unbiased estimator of $\pq$, see~\citet{Genz.Bretz2009}.
While this choice is implemented by default in the $\RprogLang$ package $\verb|anMC|$, it is not the only possible choice. The package allows for user-defined functions to estimate $\pq$. In supplementary material, Section~C we present an numerical study where the MET method introduced in~\cite{botev2016normal} is used in place of QRSVN.

In general, the estimator $\pqW$ requires two choices: $q$, the number of active dimensions, and the dimensions themselves. The decomposition of $p$ in Equation~\eqref{eq:pMaxDecom} leads to computational savings if we can approximate most of $p$ with $\pq$ for a small $q$. On the other hand a large number of active dimensions allows to intercept most of the probability mass in $p$. Here we adopt a heuristic approach to select both $q$ and $E_q$ sequentially by increasing the number of active dimensions until we meet an appropriate stopping condition. This approach, detailed in Algorithm~\ref{algo:AlgoSelq}, Appendix~\ref{sec:ActiveDims}, was chosen in the current implementation because it represents a good trade-off between speed and accuracy.

For a fixed $q$, the choice of $E_q$ plays an important role in the approximation of $p$ because it determines the error $\pqW-p$, which is always negative. Selecting $E_q$ such that $P(\max\Xq >t)$ is numerically maximized, as in~\citet{Chevalier2013}, optimally reduces the bias of $\pqW$ as an estimator for $p$. Here we are not interested in a fully fledged optimization of this quantity as the residual bias is removed with the subsequent estimation of $\Rq$, therefore, we exploit fast heuristics methods. 
The main tool used here is the excursion probability function: 
\begin{equation*}
p_t(i) = P(X_i > t) = \Phi\left(\frac{\mu_i-t}{\sqrt{\Sigma_{i,i}}}\right),
\end{equation*}
where $\Phi$ is the standard normal c.d.f. The function $p_t$ is widely used in spatial statistics \citep[see, e.g.][]{BolinLindgren2014} and Bayesian optimization \citep[see, e.g.][]{kushner1964new, Bect.etal2012}. In our setting it can be used to quickly identify the active dimensions. In fact this function takes high values at dimensions with high probability of exceeding the threshold and thus contribute the most to $p$. We propose the following methods.

\textbf{Method~A:} sample $q$ indices with probability given by $p_t$.

\textbf{Method~B:} sample $q$ indices with probability given by $p_t(1-p_t)$.

These methods require only one evaluation of the normal c.d.f. at each element of the vector $\left(\frac{\mu_{i} -t}{\sqrt{\Sigma_{i,i}}}\right)_{i=1,\dots,d}$, and are thus very fast. Both methods were already introduced for sequential evaluations of expensive to evaluate functions, see, e.g.,~\citet{Chevalier.etal2014a}.

\begin{figure}
	\centering
	\includegraphics[width=0.85\textwidth]{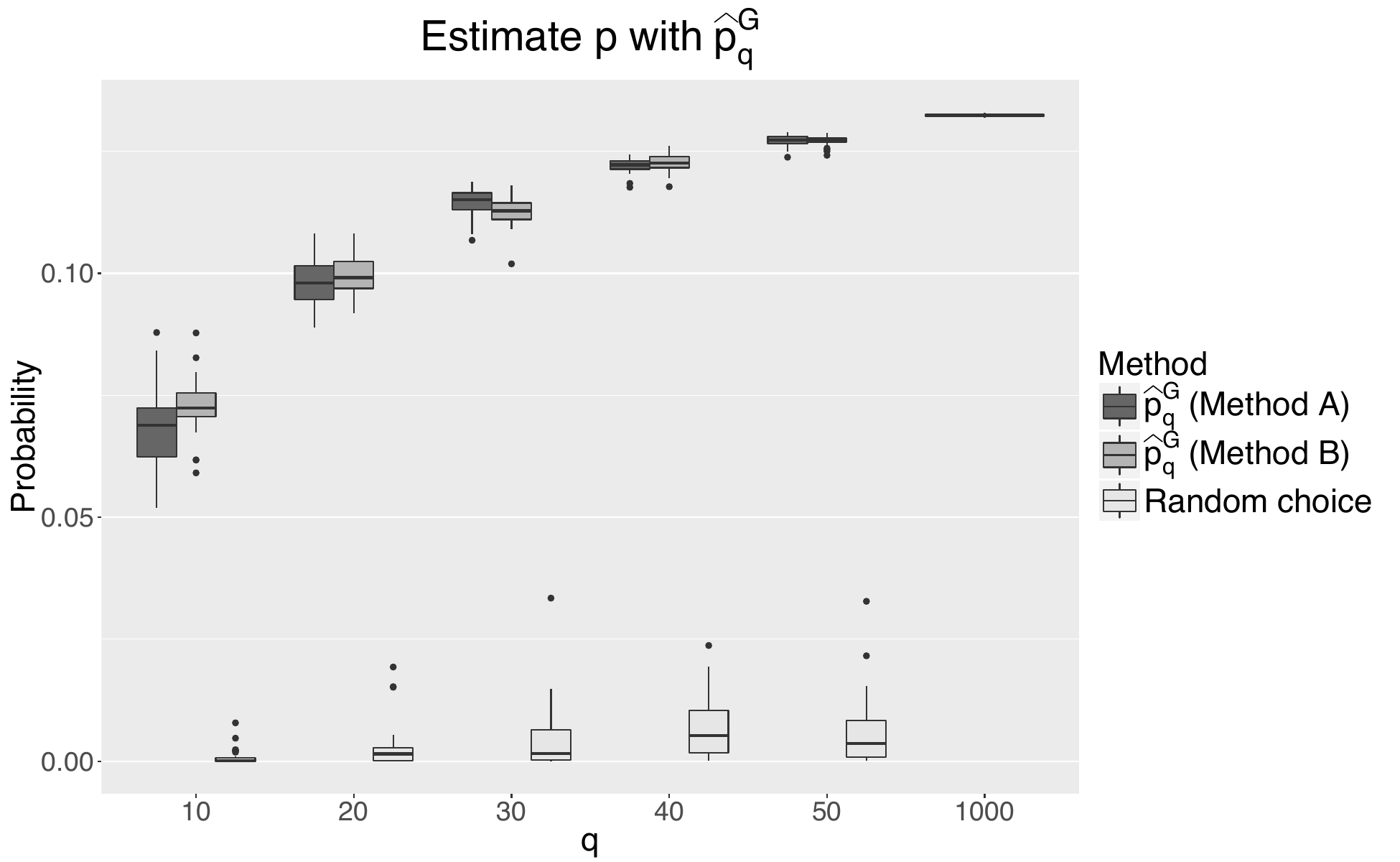}
	\caption{Distribution of $\pqG$ estimates obtained with different choices of active dimensions.}
	\label{fig:comparisonMethods}
\end{figure}

Figure~\ref{fig:comparisonMethods} shows a comparison of the estimates $\pq$ obtained with different methods to select $E_q$. We consider $30$ replications of an experiment where $\pqG$ is used to approximate $p$. The dimension of the vector $X$ is $d=1000$, the threshold is fixed at $t=11$. The vector $X$ is obtained from a discretization of a six dimensional GRF, defined on $[0,1]^6$, over the first $1000$ points of the Sobol' sequence \citep{bratley1988algorithm}. The GRF has a tensor product Mat\'ern ($\nu=5/2$) covariance kernel. We generate a non constant mean function $\muGP$ by imposing the interpolation of a deterministic function at $70$ points. The covariance kernel's hyperparameters are fixed as $\theta = [0.5,0.5,1,1,0.5,0.5]^T$ and $\sigma^2=8$, see~\citet{rasmussen2006gaussian}, Chapter~4, for details on the parametrization. In this example, the two methods clearly outperform a random choice of active dimensions.

Methods A and B work well for selecting active dimensions when the mean vector $\mu$ and the covariance matrix diagonal are anisotropic. In such cases both methods select dimensions that are a good trade-off between high variance and mean close to $t$.

The choices of $q$ and of the active dimensions influence the behaviour of the estimator for $\Rq$. This aspect is discussed in more details in the next section.

\subsection{Monte Carlo estimator for $\Rq$}
\label{subsec:MCforRq}
Debiasing $\pqW$ as an estimator of $p$ can be done at the price of estimating 
\begin{equation*}
\Rq = P \left(\max\Xminusq > t \mid \max\Xq \leq t \right).
\end{equation*}

There is no closed formula for $\Rq$, so it is approximated here via MC. 
Since $X$ is Gaussian then so are $\Xq$, $\Xminusq$ and $\Xminusq \mid \Xq = x^q$, for any deterministic vector $x^q \in \R^q$. 

In order to estimate $\Rq = P \left(\max \Xminusq > t \mid X_{i_1}\leq t, \ldots, X_{i_q}\leq t \right)$, we first generate $n$ realizations $x^q_1, \ldots, x^q_n$ of $\Xq$ such that $\Xq\leq t_q$. Second, we  compute the mean and covariance matrix of $\Xminusq$ conditional on each realization $x^q_l,\ l=1,\ldots,n$ with the following formulas
\begin{equation}
\label{eq:condMeanCov}
\mu^{-q \mid x^q_l} = \mu^{-q} + \Sigma^{-q,q} (\Sigma^{q})^{-1}(x^q_l - \mu^q), \qquad 
\Sigma^{-q \mid q} = \Sigma^{-q} - \Sigma^{-q,q}(\Sigma^q)^{-1}\Sigma^{q,-q},
\end{equation}
where $\mu^{q}, \Sigma^{q}$ and $\mu^{-q}, \Sigma^{-q}$ are the mean vector and covariance matrix of $\Xq$ and $\Xminusq$ respectively, $\Sigma^{-q,q}$ is the cross-covariance between the dimensions $E \setminus E_q$ and $E_q$, $\Sigma^{q,-q}$ is the transpose of $\Sigma^{-q,q}$. Note that the conditional covariance $\Sigma^{-q \mid q}$ does not depend on the realization $x^q_l$, therefore it can be computed before the sampling procedure. Given the mean and covariance matrix conditional on each sample $x^q_l$, we can easily draw a realization $y^{-q \mid q}_l$ from $\Xminusq \mid \Xq = x^q_l$. 
Once $n$ couples $(x^q_l, y^{-q \mid q}_l),\ l = 1, \dots, n$ are drawn from the respective distributions, an estimator for $\Rq$ is finally obtained as follows
\begin{equation*}
\RqMC = \frac{1}{n} \sum_{l=1}^n \mathbf{1}_{\max y^{-q \mid q}_l > t}.
\end{equation*}

There exists many technique to draw realizations from $\Xq$ conditional on $\Xq\leq t_q$. Here we use a crude multivariate rejection sampling algorithm \citep{robert1995simulation,horrace2005some}, however this is not the only method possible. In numerical examples in Appendix~\ref{sec:NumSmallProba} and in supplementary material we show that replacing crude rejection sampling with another sampler might be beneficial in some situations. In any case the cost of this step can be very high, in particular if we use rejection sampling then that cost is driven by the acceptance probability.
The accepted samples satisfy the condition $\Xq \leq t_q$ thus we have that the acceptance probability is $P(\Xq \leq t_q) = 1-\pq$. This shows that the choice of $q$ and of the active dimensions play an important role. If $\pq$ is much smaller than $p$, then the rejection sampler will have a high acceptance probability, however the overall method will be less efficient as most of the probability is in the remainder. On the other hand, if $q$ and the active dimensions are well chosen, the value of $\pq$ could be very close to $p$. This will also lead to a slower rejection sampler as the acceptance probability would be small. 

The second part of the procedure for $\RqMC$, drawing samples from the distribution of $\Xminusq \mid \Xq = x^q_l$, is instead less dependent on $q$ and generally less expensive than the first step. The mean vector and covariance matrix computations requires only linear algebra operations as described in Equation~\eqref{eq:condMeanCov} and
realizations of $\Xminusq \mid \Xq = x^q_l$ can be generated by sampling from a multivariate normal distribution.

The difference in computational cost between the first step and the second step of the MC procedure can be exploited to reduce the variance at a fixed computational cost. This idea is exploited by the asymmetric nested MC procedure presented in Section~\ref{sec:ANMC}. 

We denote with $\pGMC$ the unbiased estimator of $p$ defined as
\begin{equation*}
\pGMC = \pqG + (1-\pqG)\RqMC,
\end{equation*}
where $\text{GMC}$ denotes the use of Genz's method for $\pq$ and MC for $\RqW$.

\begin{figure}
	\centering
	\includegraphics[width=0.85\textwidth]{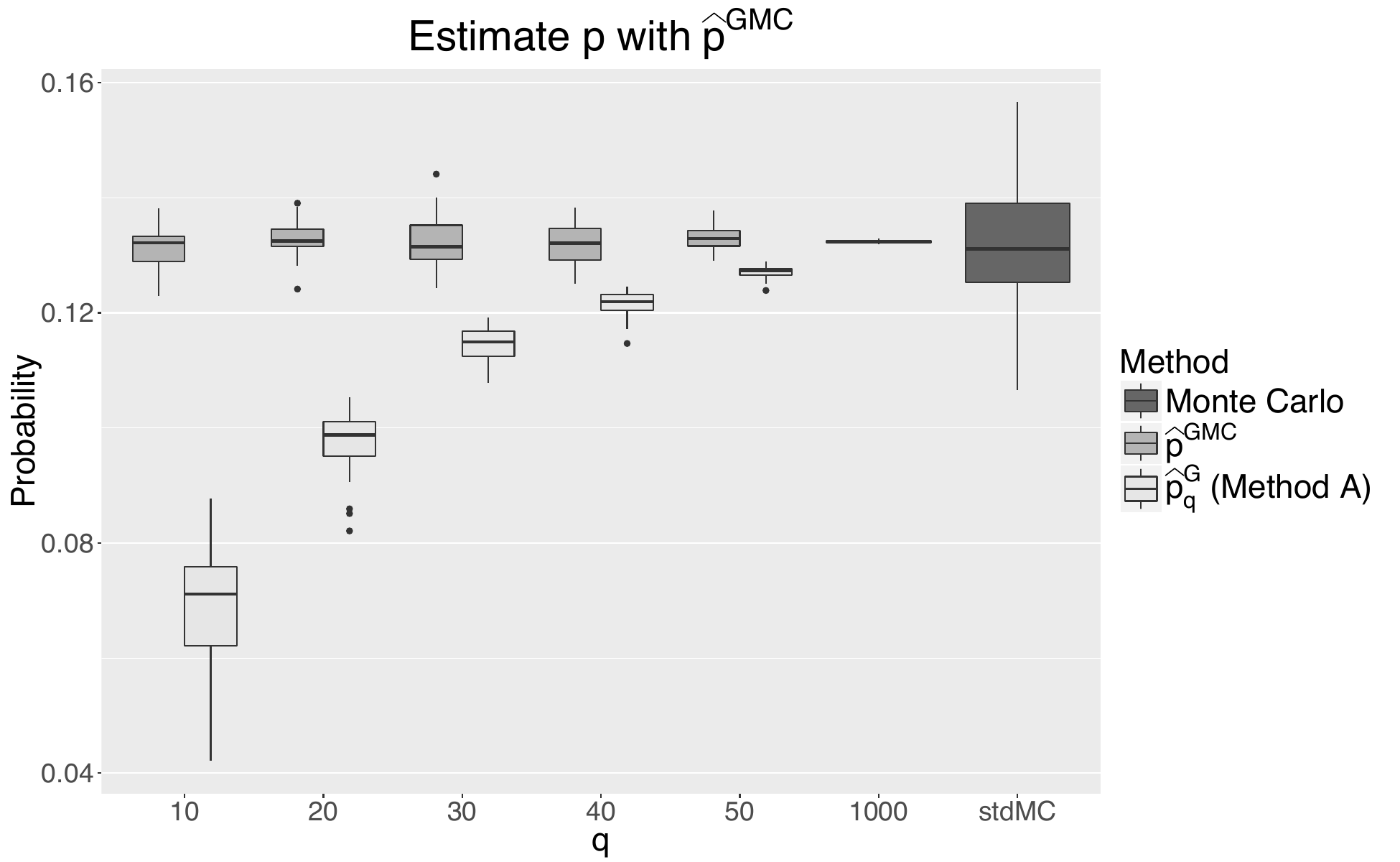}
	\caption{Estimate of $p$ with $\pGMC$ for different values of $q$. A full MC estimation of the same quantity is shown for comparison}
	\label{fig:EstimatePqRq}
\end{figure}

Figure~\ref{fig:EstimatePqRq} shows the box plots of $30$ replications of an experiment where $p$ is approximated with $\pGMC$. The set-up is the same as in Fig.~\ref{fig:comparisonMethods}. The core of the probability is approximated with $\pqG$ and the active dimensions are chosen with Method~1. The residual $\Rq$ is estimated with $\RqMC$. The remainder allows to correct the bias of $\pqG$ even with a small number of active dimensions. As comparison the results of the same experiment with a full MC estimator for $p$ are also shown. For all experiments and for each method the number of samples was chosen in order to have approximately the same computational cost. The estimator $\pGMC$ exploits an almost exact method to estimate the largest part of the probability $p$, therefore the MC estimator $\RqMC$ has less variance than a full MC procedure for a fixed computational cost.

\section{Estimation of the residual with asymmetric nested Monte Carlo}
\label{sec:ANMC}
In section~\ref{sec:MainTheory}, $\Rq$ was estimated by $\RqMC$. There exists many methods to reduce the variance of such estimators, including antithetic variables \citep{hammersley1956new}, importance sampling \citep{kahn1950random,kahn1953methods} or conditional Monte Carlo \citep{hammersley1956conditional} among many others; see, e.g.~\citet[Chapter~4]{robert2013monte}, for a broader overview. Here we focus on reducing the variance at a fixed computational cost, i.e. we are interested in increasing the estimator efficiency~\citep[Section~4.2]{lemieux2009monte}. We propose a so-called asymmetric nested Monte Carlo (anMC) estimator for $\Rq$ that increases the efficiency with a parsimonious multiple use of conditioning data. In this section we develop some useful theoretical properties of anMC estimators.

The idea is to use an asymmetric sampling scheme that assigns computational resources by taking into account the actual cost of simulating each component. A similar asymmetric sampling scheme was introduced in the particular case of comparing the performance of stopping times for a real-valued stochastic process in discrete times in~\citet{dickmannSchweizer2014}. Here we introduce this procedure in a general fashion and, in the next section, we detail it to $\RqMC$. For two measurable spaces $\mathcal{W},\mathcal{Z}$, consider two random elements $W \in \mathcal{W}$ and $Z \in \mathcal{Z}$, defined on the same probability space and not independent. We are interested in estimating the quantity 
\begin{equation}
G = \E\left[g(W,Z) \right],
\label{eq:ExpG}
\end{equation} 
where $g: \mathcal{W} \times \mathcal{Z} \rightarrow \R$ is a measurable function, assumed integrable with respect to $(W,Z)$'s probability measure. 
Let us also assume that it is possible to draw realizations from the marginal distribution of $W$, $Z$ and from the conditional distribution of $Z \mid W=w_i$, for each $w_i$ sample of $W$. 
In the spirit of a Gibbs sampler, we can then obtain realizations $(w_i,z_i)$, $i=1, \dots, n$ of $(W,Z)$ by simulating $w_i$ from the distribution of $W$ and then $z_i$ from the conditional distribution $Z \mid W=w_i$, leading to:
\begin{equation}
\widehat{G} = \frac{1}{n}\sum_{i=1}^n g(w_i,z_i).
\label{eq:wideG}
\end{equation}
This MC estimator can actually be seen as the result of a two step nested MC procedure where, for each realization $w_i$, one inner sample $z_i$ is drawn from $Z \mid W=w_i$.
Note that the estimator $\RqMC$ used in Section~\ref{sec:MainTheory} is a particular case of Equation~\eqref{eq:wideG} with $W=\Xq \mid \Xq \leq t_q$, $Z=\Xminusq$ and $g(x,y) = \mathbf{1}_{\max y > t}$. 
As noted in Section~\ref{sec:MainTheory}, drawing realizations of $\Xq \mid \Xq \leq t_q$ has a higher computational cost than simulating $\Xminusq$ because rejection sampling is required in the first case. More generally, let us denote with $C_W(n)$ the cost of $n$ realizations of $W$ and with $C_{Z \mid W}(m;w_i)$ the cost of drawing $m$ conditional simulations from $Z \mid W=w_i$. If $C_W(1)$ is much higher than $C_{Z \mid W}(1;w_i)$ then sampling several conditional realizations for a given $w_i$ 
might bring computational savings.

In the proposed asymmetric sampling scheme for each realization $w_i$ we sample $m$ realizations $z_{i,1},\dots, z_{i,m}$ from $Z \mid W= w_i$. 
Assume that we use this sampling scheme for the couples $(w_i,z_{i,j})$, $i=1, \dots, n, \ j=1,\dots,m$, then an estimator for $G$ is
\begin{equation}
\widetilde{G} = \frac{1}{nm} \sum_{i=1}^n \sum_{j=1}^m g(w_i,z_{i,j}).
\label{eq:wideTildeG}
\end{equation}

For a fixed number of samples, the estimator $\widetilde{G}$ may have a higher variance than $\widehat{G}$ due to the dependency between pairs sharing the same replicate of $W$. However, in many cases, the estimator $\widetilde{G}$ may be relevant to reduce the variance at a fixed computational time. In fact, let us fix the computational budget instead of the number of samples. If $C_{Z \mid W}(1;w_i)<C_W(1)$, then anMC may lead to an overall variance reduction thanks to an increased number of simulated pairs. In the remainder of the section, we show that, in the case of an affine cost functions, there exists an optimal number of inner simulations $m$ such that $\var(\widetilde{G}) < \var(\widehat{G})$. 
Assume
\begin{align*}
C_W(n) &= \myc_0 + \myc n \text{ and, for each sample } w_i \\
C_{Z \mid W}(m;w_i) &= C_{Z \mid W}(m) = \alpha + \beta m, 
\end{align*}
with $\myc_0, \myc, \alpha, \beta \in \R_+$ dependent on the simulators of $W$ and $Z \mid W$. 
The second equation entails that the cost of conditional simulations does not depend on the conditioning value. 
If $W=\Xq \mid \Xq \leq t_q$, $Z=\Xminusq$ as in Section~\ref{sec:MainTheory}, then $Z \mid W$ is Gaussian with mean and covariance matrix described in~\eqref{eq:condMeanCov}. In this case, the cost for sampling $Z \mid W$ is affine, with $\alpha$ describing preliminary computations and $\beta$ random number generation and algebraic operations. 
Denote with $W_1, \dots, W_n$ replications of $W$. For each $W_i$ we consider the conditional distribution $Z \mid W_i$ and $m$ replications $Z_{1,i}, \dots, Z_{m,i}$. Under these assumption the total simulation budget is 
\begin{equation*}
C_{\text{tot}}(n,m) = \myc_0 + n(\myc + \alpha + \beta m).
\end{equation*}
If the total budget is fixed, $C_{\text{tot}}(n,m) = C_{\text{fix}} \in \R_+$, then the number of replications of $W$ as a function of $m$ is 
\begin{equation*}
N_{C_{\text{fix}}}(m) = \frac{C_{\text{fix}} - \myc_0}{\myc + \alpha + \beta m}.
\end{equation*}

The following proposition shows a decomposition of $\var(\widetilde{G})$ that is useful to find the optimal number of simulations $m^*$ 
under a fixed simulation budget $C_{\text{tot}}(n,m)=C_{\text{fix}}$.

\begin{proposition}
	\label{pro:general}
	Consider $n$ independent copies $W_1, \dots, W_n$ of $W$ and, for each $W_i$, $m$ copies $Z_{i,j} = Z_j \mid W_i$ $j=1,\dots,m$, independent conditionally on $W_i$. 
	Then, 
	\begin{equation}
	\var(\widetilde{G}) = \frac{1}{n}\var(g(W_1,Z_{1,1})) - \frac{m-1}{nm} \E\big[ \var( g(W_1,Z_{1,1}) \mid W_1 )  \big].
	\label{eq:varianceDecomp}
	\end{equation}
	
\end{proposition}

\begin{corollary}
	\label{cor:OptimNum}
	Under the same assumptions, $\widetilde{G}$ has minimal variance when 
	\begin{equation*}
	m=\widetilde{m} = \sqrt{\frac{(\alpha +\myc )B}{\beta(A-B)}},
	\end{equation*}
	where $A=\var(g(W_1,Z_{1,1}))$ and $B=\E\big[ \var( g(W_1,Z_{1,1}) \mid W_1 ) \big]$. Moreover denote with $\varepsilon = \widetilde{m} - \lfloor \widetilde{m} \rfloor$, then the optimal integer is $m^* = \lfloor \widetilde{m} \rfloor$  if
	\begin{equation}
	\varepsilon < \frac{(2\widetilde{m} +1) - \sqrt{4(\widetilde{m})^2 +1}}{2}
	\label{eq:approxMstar}
	\end{equation}
	or $m^* = \lceil \widetilde{m} \rceil$ otherwise.
\end{corollary}


\begin{proposition}
	\label{pro:comparisonVar}
	Under the same assumptions, if
	$m^* > \frac{2(\alpha +\myc)B}{(\myc+\alpha)B+ \beta(A-B)}$
	then $\var(\widetilde{G}) = \var(\widehat{G}) \left[1- \eta\right]$,
	where $\eta \in (0,1)$.
	
\end{proposition}

\subsection{Algorithmic considerations}
\label{subsec:AlgorithmANMC}

In order to compute $m^*$, we need the quantities $A=\var( g(W_1,Z_{1,1}) )$ and $B=\E\big[ \var( g(W_1,Z_{1,1}) \mid W_1 ) \big]$ and the constants $\myc_0$, $\myc$, $\alpha$ and $\beta$. 
$A$ and $B$ depend on the specific problem at hand and are usually not known in advance. Part of the total computational budget is then needed to estimate $A$ and $B$. 
This preliminary phase is also used to estimate the system dependent constants $\myc$ and $\beta$. 
Algorithm~\ref{algo:Algo1} reports the pseudo-code for anMC. 


\begin{algorithm}
	\SetKwInOut{Input}{Input}\SetKwInOut{Output}{Output}
	\Input{$\mu_W, \mu_Z, \Sigma_W, \Sigma_Z, \Sigma_{WZ}, g, C_{\text{tot}}$}
	\Output{$\widetilde{G}$}	
	\nlset{Part 0: \Indm}  estimate $\myc_0, \myc, \beta, \alpha$ \;
	\nlset{initialize} 
	compute the conditional covariance $\Sigma_{Z \mid W}$ and initialize $n_0, m_0$\;
	\nlset{Part 1: \Indm} \For{$i \leftarrow 1$ \KwTo $n_0$}{
		\nlset{estimate $A,B$}  	simulate $w_i$ from the distribution of $W$ and compute $\mu_{Z \mid W=w_i}$\;
		draw $m_0$ simulations $z_{i,1}, \dots, z_{i,m_0}$ from the conditional distribution $Z \mid W=w_i$\;
		estimate $\E\left[ g(W,Z) \mid W= w_i  \right]$ with $\tilde{E}_i = \frac{1}{m_0}\sum_{j=1}^{m_0} g(w_i,z_{i,j})$\;
		estimate $\var\left( g(W,Z) \mid W= w_i  \right)$ with $\tilde{V}_i = \frac{1}{m_0-1}\sum_{j=1}^{m_0} (g(w_i,z_{i,j})-\tilde{E}_i)^2$\;
	}
	compute $\widetilde{m} = \sqrt{\frac{(\alpha+\myc) \frac{1}{n_0} \sum_{i=1}^{n_0}{\tilde{V}_i}}{\beta\frac{1}{n_0-1}\sum_{i=1}^{n_0}(\tilde{E}_i - \frac{1}{n_0}\sum_{i=1}^{n_0} \tilde{E}_i)^2 }}$, $m^*$ as in Corollary~\ref{cor:OptimNum} and $n^*=N_{C_{\text{fix}}}(m^*)$\;
	
	\nlset{Part 2: \Indm}  \For{$i \leftarrow 1$ \KwTo $n^*$}{
		\nlset{compute $\widetilde{G}$}	\eIf{$i\leq n_0$}{
			\For{$j \leftarrow 1$ \KwTo $m^*$}{
				\eIf{$j \leq m_0$}{
					use previously calculated $\widetilde{E}_i$ and $\widetilde{V}_i$\;
				}{
				simulate $z_{i,j}$ from the distribution $Z \mid W=w_i$\;
				compute $\widetilde{E}_i = \frac{1}{m^*}\sum_{j=1}^{m^*} g(w_i,z_{i,j})$\;
			}
		} 
	}{
	simulate $w_i$ from the distribution of $W$ and compute $\mu_{Z \mid W=w_i}$\;
	\For{$j \leftarrow 1$ \KwTo $m^*$}{
		simulate $z_{i,j}$ from the conditional distribution $Z \mid W=w_i$\;
	}
	compute $\widetilde{E}_i = \frac{1}{m^*}\sum_{j=1}^{m^*} g(w_i,z_{i,j})$\;			
}

}

estimate $\E\left[ g(W,Z) \right]$ with $\widetilde{G} = \frac{1}{n^*}\sum_{i=1}^{n^*} \tilde{E}_i$\;

\caption{Asymmetric nested Monte Carlo.}
\label{algo:Algo1}
\end{algorithm}

\subsection{Estimate $p$ with $\pGANMC$}
The anMC algorithm can be used to reduce the variance compared to $\Rq$'s MC estimate proposed in Section~\ref{subsec:MCforRq}. In fact, let us consider $W=\Xq \mid \Xq \leq t_q$ and $Z=\Xminusq$. We have that $W$ is expensive to simulate as it requires sampling from a truncated normal while, for a given sample $w_i$, $Z \mid W=w_i$ is Gaussian with mean and covariance matrix described in Equation~\eqref{eq:condMeanCov}. It is generally much cheaper to obtain samples from $Z \mid W=w_i$ than from $W$. Moreover, as noted earlier, $\Rq$ can be written in the form of Equation~\eqref{eq:ExpG} with $g(x,y) = \mathbf{1}_{\max y > t}$. By following Algorithm~\ref{algo:Algo1} we calculate $m^*$, sample $n^*$ realizations $w_1, \dots, w_{n^*}$ of $W$ and for each realization $w_i$ obtain $m^*$ samples $z_{i,1}, \dots, z_{i,m^*}$ of $Z \mid W = w_i$. We estimate $\Rq$ via 
\begin{equation*}
\RqANMC = \frac{1}{n^* m^*} \sum_{i=1}^{n^*}\sum_{j=1}^{m^*}  \mathbf{1}_{\max z_{i,j} > t}.
\end{equation*}

Finally plugging in $\RqANMC$ and $\pqG$ in Equation~\eqref{eq:pMaxDecom}, we obtain 
\begin{equation*}
\pGANMC = \pqG + (1-\pqG)\RqANMC.
\end{equation*}

\begin{figure}
	\begin{subfigure}{0.5\textwidth}
		\centering
		\includegraphics[width=\textwidth]{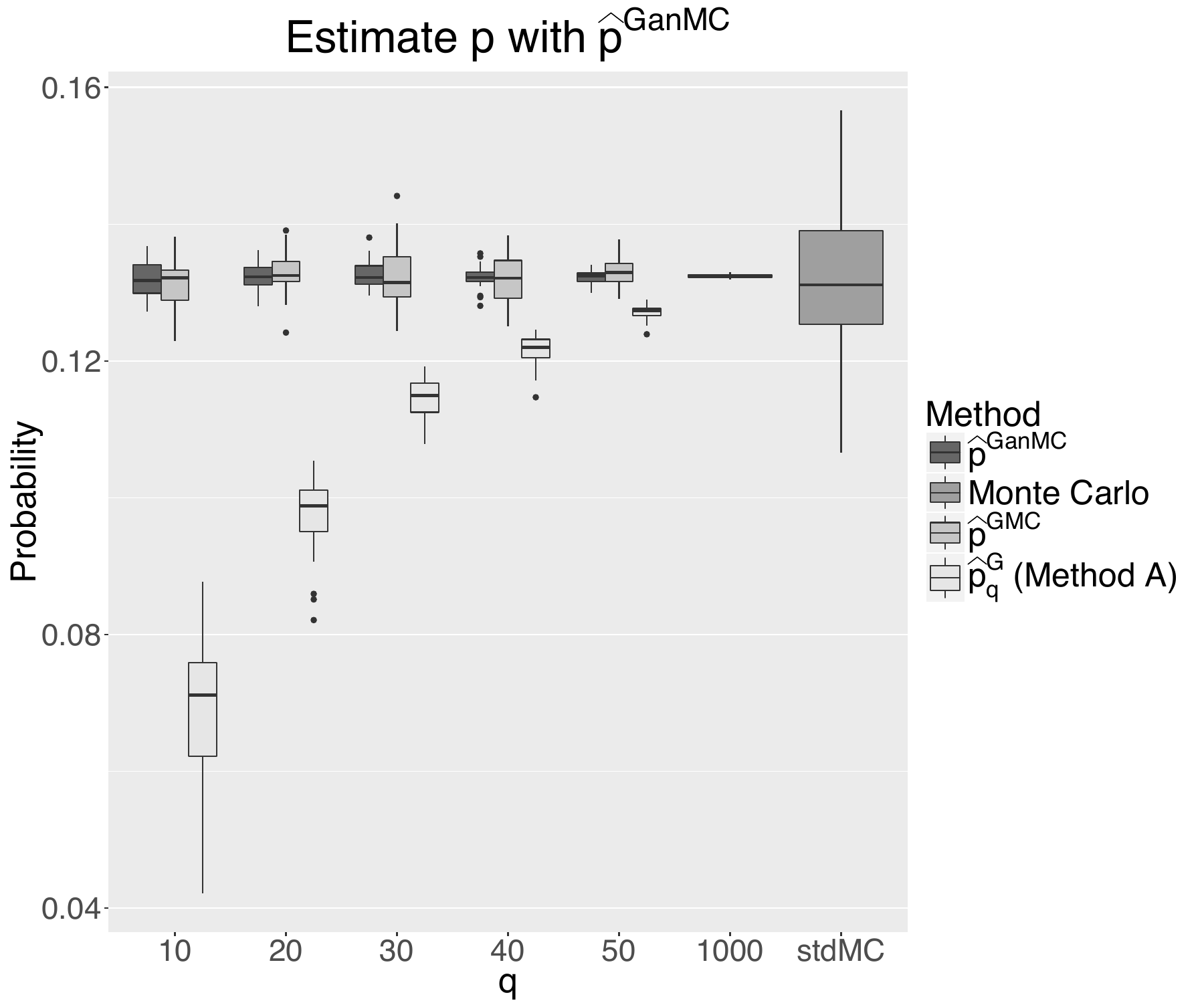}
		\caption{Probability values.}
		\label{fig:EstimatePganmc}
	\end{subfigure} \hfill
	\begin{subfigure}{0.5\textwidth}
		\centering
		\includegraphics[width=\textwidth]{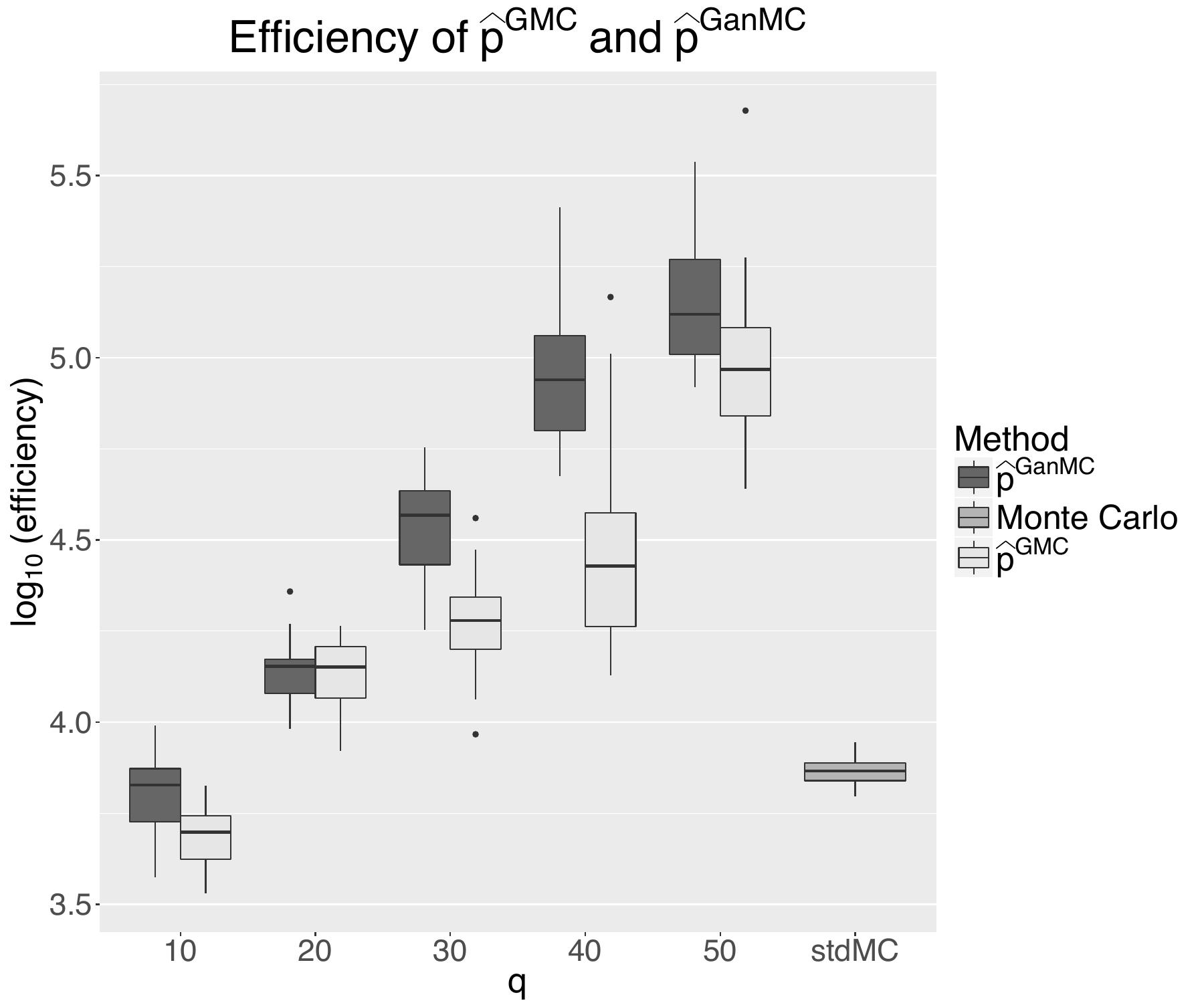}
		\caption{Efficiency. Values in logarithmic scale.}
		\label{fig:EfficiencyMC}
	\end{subfigure}
	\caption{Comparison of results with $\pqG$, $\pGMC$, $\pGANMC$ and standard MC on $30$ replications of the example introduced in Fig.~\ref{fig:comparisonMethods}.}
	\label{fig:CompMCprobEfficiency}
\end{figure}

Figure~\ref{fig:EstimatePganmc} shows a comparison of results using $30$ replications of the experiment presented in Section~\ref{subsec:MCforRq}. 
Results obtained with a MC estimator are shown for comparison.

While the simulations of all experiments were obtained under the constraint of a fixed computational cost, the actual time to obtain the simulations was not exactly the same. In order to be able compare the methods in more general settings we further rely on the notion of efficiency. For an estimator $\widehat{p}$, we define the efficiency~\citep[Section~4.2]{lemieux2009monte} as
\begin{equation}
\operatorname{Eff}[\widehat{p}] = \frac{1}{\var(\widehat{p})\operatorname{time}[\widehat{p}]},
\label{eq:efficiency}
\end{equation}
where $\operatorname{time}[\widehat{p}]$ denotes the computational time of the estimator $\widehat{p}$. 


Figure~\ref{fig:EfficiencyMC} shows a comparison of the efficiency of $\pGMC$ and $\pGANMC$ with a full Monte Carlo estimator. With as few as $q=50$ active dimensions we obtain an increase in efficiency of around $10$ times on average over the $30$ replications of the experiment with the estimator $\pGMC$. The estimator $\pGANMC$ shows a higher median efficiency than the others for all $q \geq 20$.

\section{Numerical studies}
\label{sec:NumStudies}

\subsection{Choice of the number of inner samples}
\label{subsec:NumInner}

In this section we study the efficiency of the anMC method compared with a standard MC method for different choices of $m$. Here we do not select the optimal $m^*$ defined in Corollary~\ref{cor:OptimNum}, but we study the efficiency as a function of $m$. In many practical situations even if part~1 of Algorithm~\ref{algo:Algo1} does not render the optimal $m^*$ the anMC algorithm is still more efficient than a standard MC if the chosen $m$ is close to $m^*$.

We consider a similar setup to the experiment presented in Section~\ref{subsec:pq}. Here we start from a GRF with tensor product Mat\'ern ($\nu=5/2$) and a non constant mean function $\muGP$ different from the example in Section~\ref{subsec:pq}, initialized as conditional mean on $60$ randomly generated values at a fixed design on $[0,1]^6$. The hyperparameters are fixed as $\theta=[0.5,0.5,1,1,0.5,0.5]^T$ and $\sigma^2=8$. The GRF is then discretized over the first $d=1000$ points of the Sobol sequence to obtain the vector $X$. We are interested in $1-p=P(X < t)$, with $t=5$. We proceed by estimating $p$ with $\pGMC$ and $\pGANMC$ for different choices of $m$ to compare their efficiency. The initial part $\pq$ is computed once with estimator $\pqG$ with $q$ and the active dimensions chosen with Algorithm~\ref{algo:AlgoSelq}, Method~B. The number of outer simulations in the anMC algorithm is kept fixed to $n=10,000$ and we only vary $m$. For each $m$, the anMC estimation is replicated $20$ times.

\begin{figure}[t]
	\begin{subfigure}{0.5\textwidth}
		\centering
		\includegraphics[width=\linewidth]{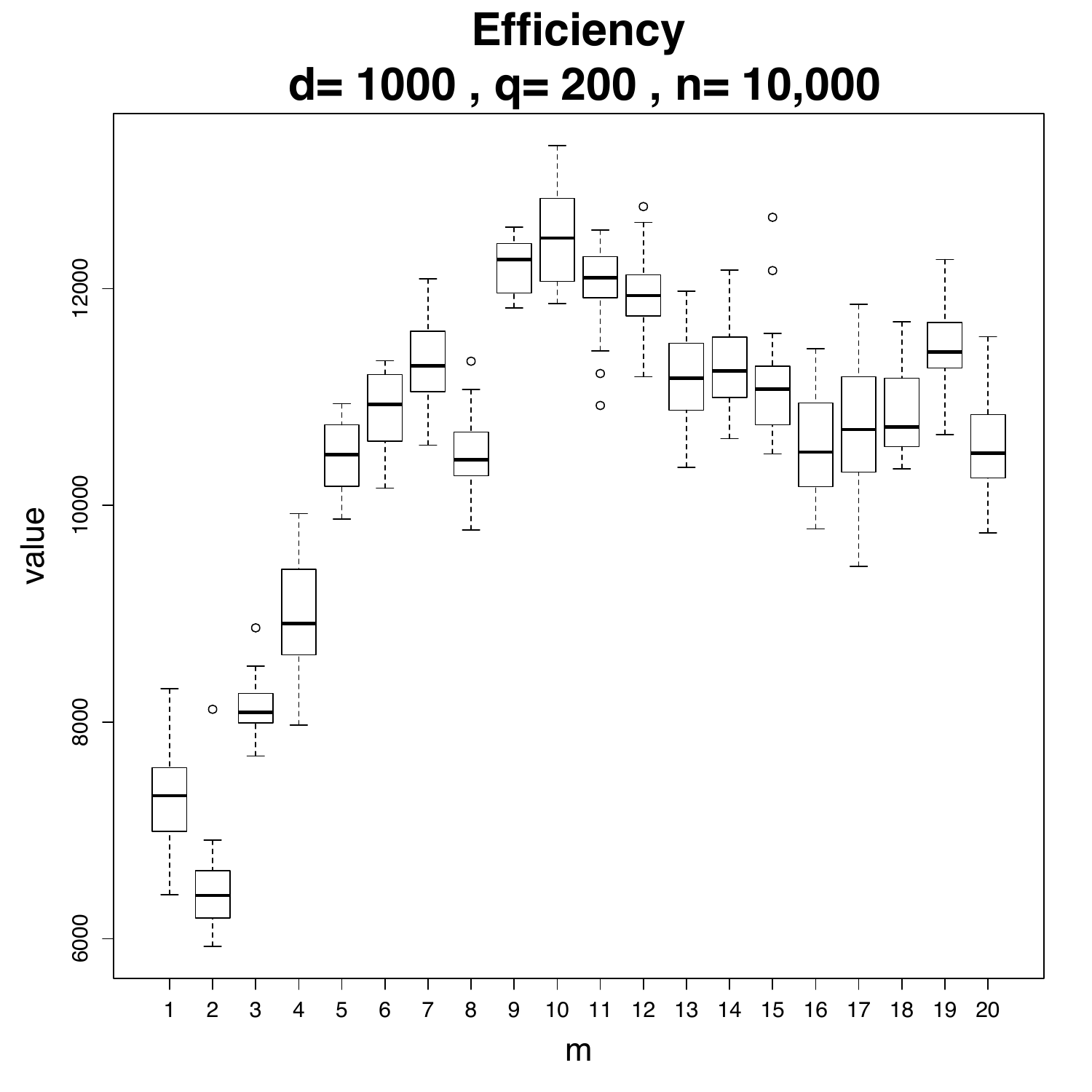}
		\caption{High probability state, $t=5$, $\pGANMC=0.9644$.}
		\label{fig:Eff6DThresh5d1000}
	\end{subfigure} \hfill
	\begin{subfigure}{0.5\textwidth}
		\centering
		\includegraphics[width=\linewidth]{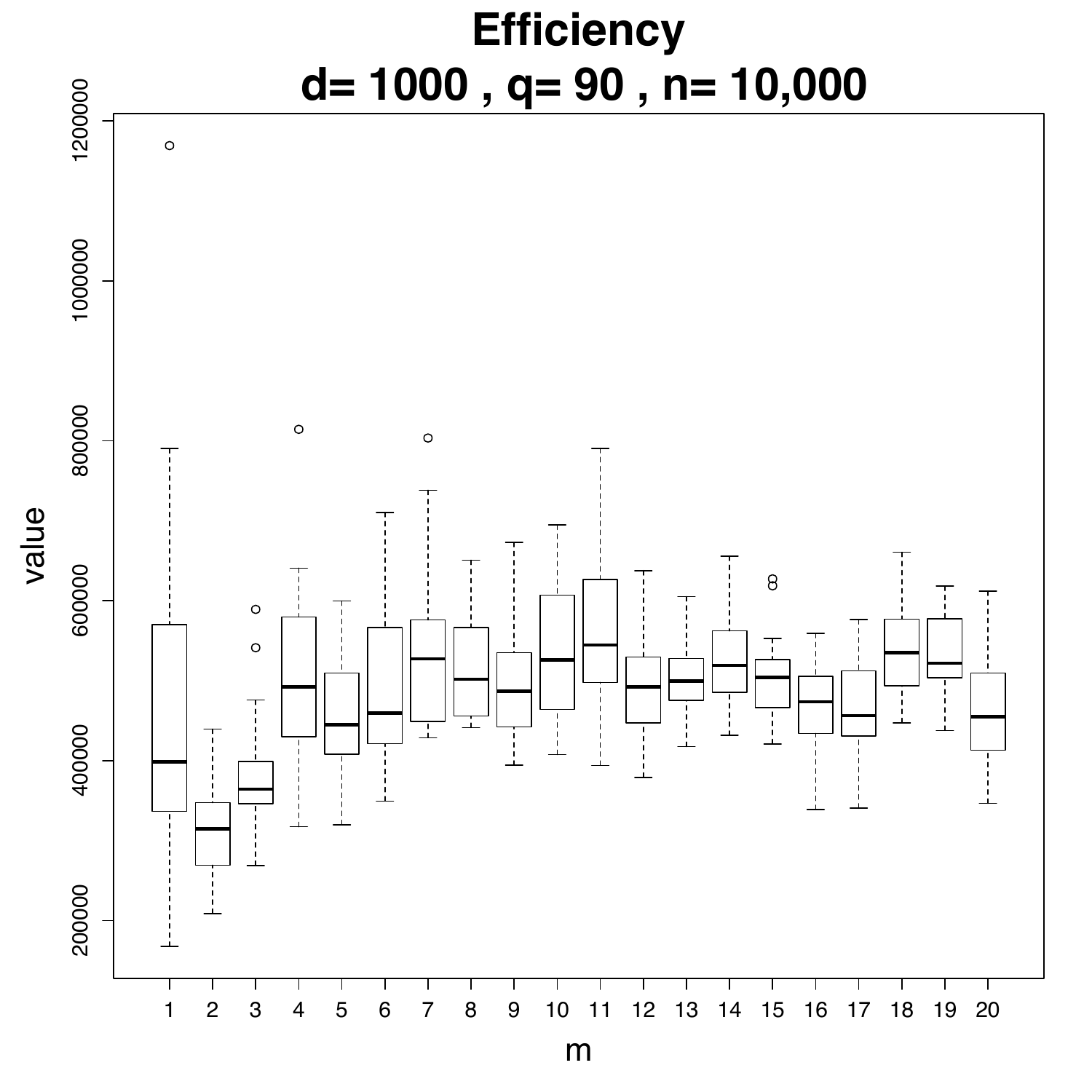}
		\caption{Low probability state, $t=7.5$, $\pGANMC=0.1178$.}
		\label{fig:Eff6DThresh7_5d1000}
	\end{subfigure}
	\caption{Efficiency of $\pGANMC$ estimator versus the number of inner simulations $m$. For each $m$ the experiment is reproduced $30$ times.}
	\label{fig:Eff6DvsM}
\end{figure}

The median estimated value for $p$ is $\widehat{p}= 0.9644$. Most of the probability is estimated with $\pq$, in fact $\pqG= 0.9636$. Figure~\ref{fig:Eff6DThresh5d1000} shows $\operatorname{Eff}[\widehat{p}]$ computed with the overall variance of $\widehat{p}$. A choice of $m=10$ leads to a median increase in efficiency of $73\%$ compared to the MC case. In this example, both the probability to be estimated and $\pq$ are close to $1$, thus the acceptance probability for $\RqW$ is low. In this situation the anMC method is able to exploit the difference in computational costs to provide a more efficient estimator for $\RqW$. 

In order to study the effect of the acceptance probability on the method's efficiency we change the threshold in the previous example to $t=7.5$ by keeping the remaining parameters fixed. The value of $p$ is smaller, $\widehat{p}=0.1178$. The number of active dimensions $q$, chosen with Algorithm~\ref{algo:AlgoSelq}, is smaller ($q=90$) as the probability mass is smaller. The value of $\pq$ ($\pqG=0.1172$) is much smaller than in the previous case and this leads to a higher acceptance probability for $\RqW$. Figure~\ref{fig:Eff6DThresh7_5d1000} shows efficiency of the method as a function of $m$. Here the anMC method does not bring significant gains over the MC method as the the ratio between the cost of rejection sampling and the conditional simulations in $\RqW$ is close to one. The estimated $m^*$ is equal to $1.91$, thus it is smaller than the minimum threshold of Proposition~\ref{pro:comparisonVar} that guarantees a more efficient anMC algorithm.

\subsection{Comparison with state of the art}
\label{subsec:CompSoA}

In this section we compare the GMC and GanMC methods, with the default options as implemented in the $\RprogLang$~package \verb|anMC|, with available state-of-the-art algorithms to estimate $\pi(t)$. In particular, we compare this implementation with: 
\begin{description}
	\item[QRSVN] an implementation of Genz method \citep{Genz.Bretz2009} in the $\RprogLang$~package \verb|mvtnorm|~\citep{Genz.etal_mvtnorm}, function~\verb|pmvnorm|;
	\item[GHK] an implementation of GHK method \citep{geweke1991efficient,hajivassiliou1998method} in the $\RprogLang$~package \verb|bayesm|~\citep{Rossi_bayesm}, function~\verb|ghkvec|;
	\item[MET] $\RprogLang$~implementation of the minimax-exponentially-tilted (MET) method \citep{botev2016normal} in the package \verb|TruncatedNormal|~\citep{Botev_TN}, function~\verb|mvNcdf|;.
\end{description}

We consider the example introduced in Section~\ref{subsec:NumInner} and we increase the dimension of the problem $d$ by considering finer discretizations of the underlying GRF. For example, the vector $X$ of dimension $d=100$ is obtained from the GRF discretized on the first $100$ points of the $6$-dimensional Sobol' sequence. As the dimension $d$ increases the probability $\pi(t)$ changes, thus providing different setups. Each experiment is replicated $15$ times.

\begin{figure}
		\centering
		\includegraphics[width=0.5\linewidth]{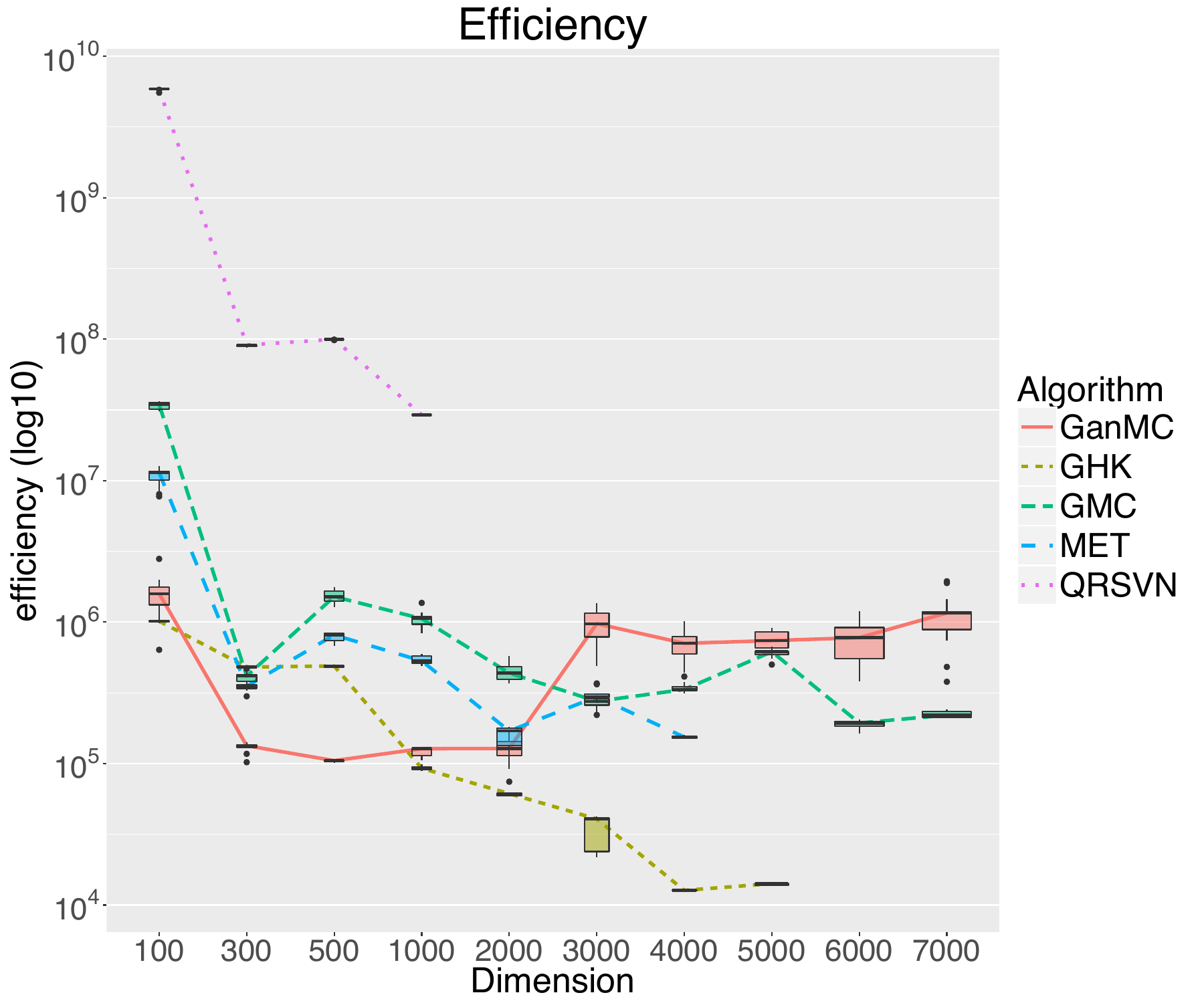}
	\caption{Efficiency of the probability estimator for $\pi(t)$, with $t=5$, versus the dimension $d$. For each $d$ the experiment is reproduced $15$ times. Values in logarithmic scale. The median estimated value for $p=1-\pi(t)$ ranges from $0.33845$ to $0.99876$.}
	\label{fig:CompEff6DThresh5}
\end{figure}

Figure~\ref{fig:CompEff6DThresh5} presents a comparison of the estimator's efficiency for the problem of computing $\pi(t)$, with $t=5$. In this setup the value of $\pi(t)$ varies between $0.66155$ for $d=100$ and $0.00124$ for $d=7000$. The most efficient algorithm is the QRSVN Genz method, however this implementation does not scale to dimensions higher than $1000$. The GMC algorithm is the second most efficient in all dimensions except $d=2000$ where it is the most efficient. The GanMC algorithm is instead the most efficient when $d$ is greater than $2000$. This effect is explained by the efficiency gains brought by $\RqANMC$ when the rejection sampler is expensive. If $d >2000$, the probability $P(\Xq \leq t_q)$ is always smaller than $0.01$, thus the rejection sampler becomes much more expensive than the conditional sampler in the estimation of the remainder $\RqW$. Algorithms GHK and MET allowed estimates until dimension $d=5000$ and $d=4000$ respectively before running in memory overflows. The GanMC algorithm is $45$ times more efficient than the GHK algorithm for $d=5000$ and $3.8$ times more efficient than MET for $d=4000$. It is also $5$ times more efficient than GMC for $d=7000$. Computations were carried on the cluster of the University of Bern on machines with Intel Xeon CPU 2.40GHz and 16 GB RAM. 

Appendix~\ref{sec:NumSmallProba} shows a comparison of the efficiency for $t=7.5$ and for $t=3$. By changing the level $t$ we obtain radically different situations: in the first case the acceptance probability of the rejection sampler becomes quite high, thus limiting the benefit of the anMC procedure. In the second case $\pi(t)$ becomes very small as the dimension increases, thus making the problem of estimating $\Rq$ out of reach with rejection sampling. In this example we present an alternative to the default choice implemented in \verb|anMC|.

\section{Application: efficient computation of conservative estimates}
\label{sec:ConservativeEsts}

A problem where the anMC method leads to substantial increases in efficiency is conservative excursion set estimation relying on Gaussian field models. 
We consider an expensive to evaluate system described by a continuous function $f: D \subset \R^\ell \rightarrow \R, \ell\geq 1$, where $D$ is a compact domain, and 
we focus on estimating, for some fixed threshold $t \in \R$, the set 
\begin{equation*}
\Gamma^* = \{x \in D : f(x) \leq t \}. 
\end{equation*}
Such problems arise in many applications such as reliability engineering (see, e.g., \cite{picheny2013quantile}, \cite{Chevalier.etal2014}) climatological studies \citep{BolinLindgren2014,frenchSain2013spatio} or in natural sciences \citep{bayarri.etal2009using}. Often $f$ is seen as expensive to evaluate black-box \citep{Sacks.etal1989} and can only be evaluated with computer simulations. We assume here that $f$ was only evaluated at points $\chi_k = \{x_1, \dots, x_k\} \subset D$ and the associated responses are denoted with $f(\chi_k)= \left(f(x_1), \dots, f(x_k)\right) \in \R^k$ and we are interested in giving an estimate of $\Gamma^*$ starting from these $k$ evaluations. 

In a Bayesian framework we consider $f$ as a realization of a GRF $(\xi_x)_{x \in D}$ with prior mean function $\mathfrak{m}$ and covariance kernel $\mathfrak{K}$. 
A prior distribution of the excursion set is hence obtained by thresholding $\xi$, thus obtaining the following random closed set
\begin{equation*}
\Gamma = \{ x\in D: \xi_x \leq t\}.
\end{equation*}
Denoting with $\xi_{\chi_k}$ the random vector $(\xi_{x_1}, \dots, \xi_{x_k})$, we can then condition $\xi$ on the observations $f(\chi_k)$ and obtain a posterior distribution for the field $\xi_x \mid \xi_{\chi_k} = f(\chi_k)$. This gives rise to a posterior distribution for $\Gamma$. 
Different definitions of random closed set expectation (\citet{Molchanov2005}, Chapter~2) can be used to summarize this posterior distribution and to provide estimates for $\Gamma^*$. In \citet{Chevalier.etal2013b}, for example, the Vorob'ev expectation was introduced in this setting. Let us briefly recall this definition. We denote with $p_{\Gamma,k}: D \rightarrow [0,1]$ the coverage function of the posterior set $\Gamma \mid \xi_{\chi_k} = f(\chi_k)$, defined as
\begin{equation*}
p_{\Gamma,k}(x) = P_k(x \in \Gamma), \ x \in D,
\end{equation*}
where $P_k(\cdot) = P(\cdot \mid \xi_{\chi_k} = f(\chi_k))$. This function associates to each point in $D$ its probability of being inside the posterior excursion set. The function $p_{\Gamma,k}$ gives rise to a family of excursion set estimates: for each $\rho \in [0,1]$ we can define the posterior $\rho$-level Vorob'ev quantile of $\Gamma$ 
\begin{equation*}
Q_{\rho}= \{x \in D : p_{\Gamma,k}(x) \geq \rho \}.
\end{equation*}
The Vorob'ev expectation of $\Gamma$ \citep{Molchanov2005} is the quantile $Q_{\rho_{V}}$ that satisfies $\lvert Q_\rho \rvert \leq \E_k[\lvert \Gamma \rvert] \leq \lvert Q_{\rho_{V}} \rvert$ for all $\rho \geq \rho_{V}$, where $\lvert A \rvert$ denotes the volume of a set $A \subset \R^l$.
The set $Q_{\rho_V}$ consists of the points that have high enough marginal probability of being inside the excursion set. 
In some applications, however, it is important to provide confidence statements on the whole set estimate. Conservative estimates introduced in \citet{BolinLindgren2014} for Gaussian Markov random fields address this issue. A conservative estimate of $\Gamma^*$ is 
\begin{equation}
C_{\Gamma,k} = \underset{C \subset D}{\arg\max} \{ \lvert C\rvert  : P_k(C \subset \{\xi_x \leq t\}) \geq \alpha\},
\label{eq:ConsEstDef}
\end{equation}
where $\lvert C \rvert$ denotes the volume of $C$.  

The estimation of the object in Equation~\eqref{eq:ConsEstDef}, however, leads to major computational issues. First of all we need to select a family of sets to use for the optimization procedure in Equation~\eqref{eq:ConsEstDef}. Here we follow \citet{BolinLindgren2014} and select the Vorob'ev quantiles as family of sets. This family has the advantage that it is parametrized by one real number $\rho$ and thus it renders the optimization straightforward. Algorithm~\ref{algo:AlgoConsEst} details the optimization procedure.	

\begin{algorithm}[!t]
	\SetKwInOut{Input}{Input}\SetKwInOut{Output}{Output}
	\Input{\vspace{-0.1cm}
$\mathfrak{m}_k,\mathfrak{K}_k$, conditional mean and covariance of $\xi \mid \xi_{\chi_k} = f(\chi_k)$, and \\ $G$, fine discretization design;
\vspace{-0.2cm}}
	\Output{Conservative estimate for $\Gamma^*$ at level $\alpha$.\vspace{0.1cm}}	
	\nlset{Part 0: \Indm} sort the points in $G$ in decreasing order of $p_{\Gamma,k}$, with indices $G_s = \{i_1, \dots i_m\}$\; 
	\nlset{compute $i_B$, $i_T$} find the highest index $i_T$ such that $\prod_{j=1}^{T}p_{\Gamma,k}(G_s)[i_j] \geq \alpha$\;
	find the highest index $i_B$ such that $p_{\Gamma,k}(G_s)[i_B] \geq \alpha$\;
	evaluate mean and covariance matrix $\mathfrak{m}_k(i_B)$ and $\Sigma_{i_B,i_B}$\; 
	\nlset{Part 1: \Indm} initialize $i_L = i_T$, $i_R=i_B$   \;
	\nlset{Initialize dichotomy}
	estimate $P_L= P_k(Q_{\rho_{i_L}} \subset \{\xi_x \leq t\})$, $P_R = P_k(Q_{\rho_{i_R}} \subset \{\xi_x \leq t\})$ with GanMC \;
	\nlset{Part 2: \Indm} \While{$P_R < \alpha$ and $( i_R-i_L) \geq 2$}{
		\nlset{optimization} next evaluation $i_{\text{next}} = \frac{i_L+i_R}{2}$\;
		estimate $P_{\text{next}} = P_k(Q_{\rho_{i_\text{next}}} \subset \{\xi_x \leq t\})$ with GanMC\;
		\eIf{$P_{\text{next}} \geq \alpha$}{
			$i_L =i_{\text{next}}$, $i_R =i_R$\;
		}{	$i_L =i_L$, $i_R =i_{\text{next}}$\;}
	}
	
	\caption{Conservative estimates algorithm.}
	\label{algo:AlgoConsEst}
\end{algorithm}

Second, for each candidate $Q$ we need to evaluate $P_{\text{next}} = P_k(Q \subset \{\xi_x \leq t\})$, the probability that $Q$ 
is inside the excursion. In fact, this quantity is a high dimensional orthant probability. For a Vorob'ev quantile $Q_{\rho_\prime}$, discretized over the points $c_1, \dots, c_r$,  
\begin{equation*}
P_k(Q_{\rho^\prime} \subset \{\xi_x \leq t\}) = P_k(\xi_{c_1} \leq t, \dots, \xi_{c_{r}} \leq t) = 1- P_k(\underset{i=1,\dots, r}{\max} \xi_{c_i} >t ).
\end{equation*}
Thus we use the estimator $\pGANMC$ to approximate $1-P_k(Q_{\rho^\prime} \subset \{\xi_x \leq t\})$. 
The use of anMC allows resolutions for the discretized Vorob'ev quantiles that seem out of reach otherwise.

\begin{figure}
	\centering
	\begin{subfigure}[b]{0.475\textwidth}
		\includegraphics[width=\linewidth]{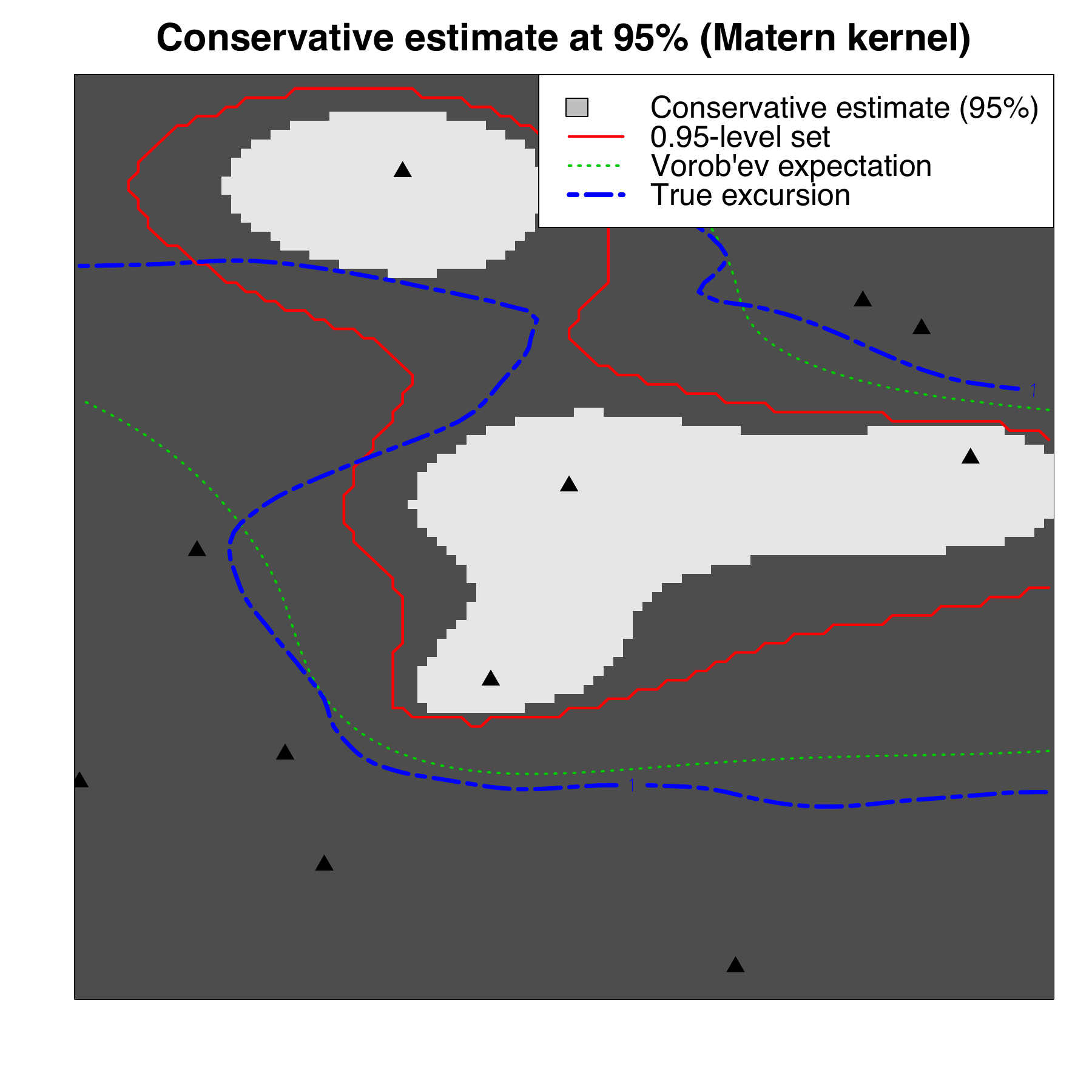}
		\caption{Realization obtained with a Mat\'ern kernel.}
		\label{fig:ConsEstimateMatern}
	\end{subfigure}
	\hfill
	\begin{subfigure}[b]{0.475\textwidth}
		\includegraphics[width=\textwidth]{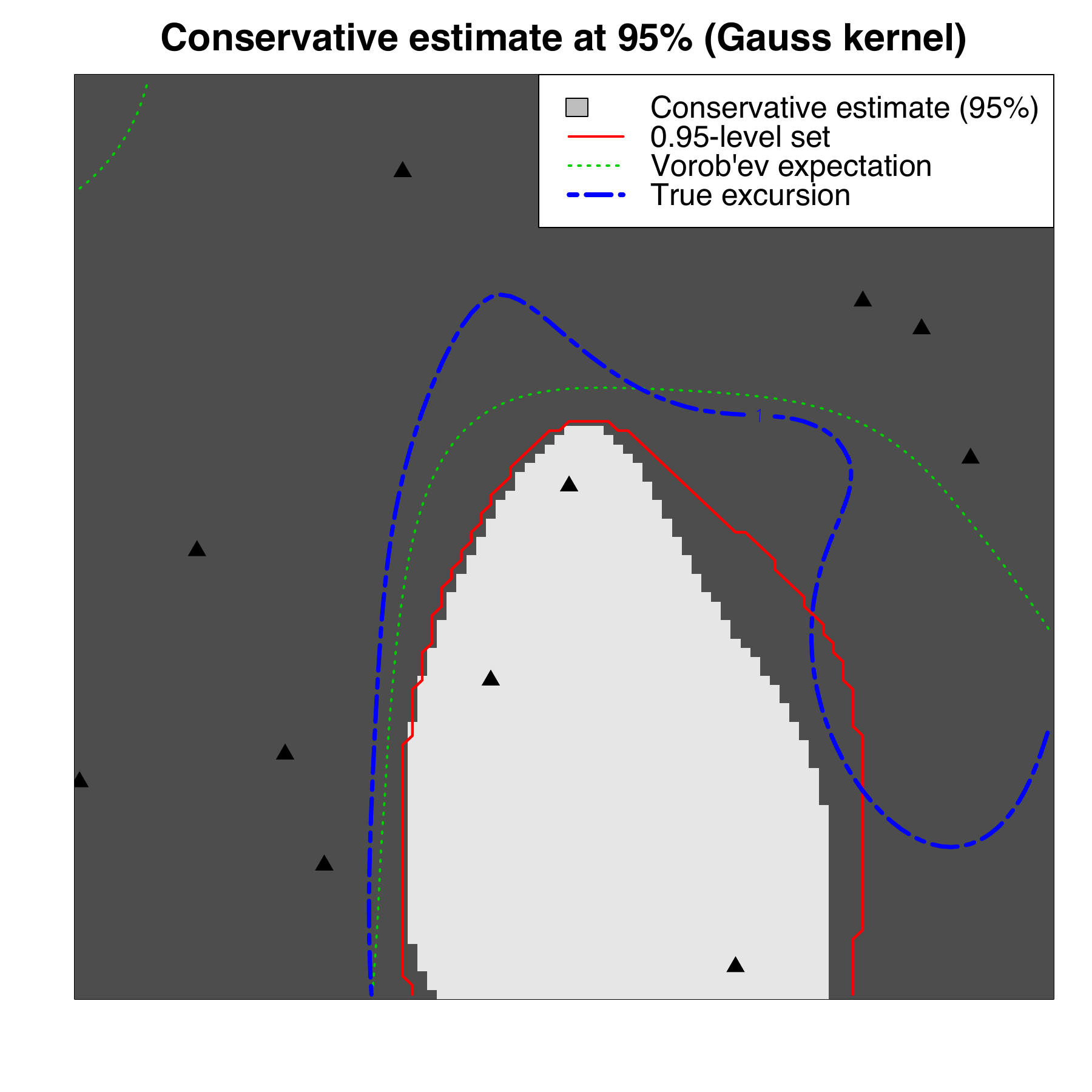}
		\subcaption{Realization obtained with Gaussian kernel.}
		\label{fig:ConsEstimateGauss}
	\end{subfigure}
	\caption{Conservative estimates at $95\%$ (white region) for the excursion below $t=1$. Both models are based on $15$ evaluations of the function (black triangles). The true excursion level is plotted in blue, the Vorob'ev expectation in green and the $0.95$-level set in red.}
	\label{fig:ConsEstimates}
\end{figure}

We apply Algorithm~\ref{algo:AlgoConsEst} to a two dimensional artificial test case. We consider as function $f$ a realization of a GRF $(\xi_x)_{x \in D}$, where $D \subset \R^2$ is the unit square. We consider two parametrizations for the prior covariance kernel: a tensor product Mat\'ern covariance kernel with $\nu=5/2$, variance $\sigma^2=0.5$ and range parameters  $\mathbf{\theta}=[0.4,0.2]$ and a Gaussian covariance kernel with variance $\sigma^2=0.5$ and range parameters  $\mathbf{\theta}=[0.2,0.4]$. In both cases we assume a prior constant mean function. We are interested in the set $\Gamma^*$ with $t=1$. For both cases we consider $k=15$ evaluations of $f$ at the same points chosen by Latin hypercube sampling. Figures~\ref{fig:ConsEstimateMatern} and~\ref{fig:ConsEstimateGauss} show the conservative estimate at level $95\%$ compared with the true excursion, the Vorob'ev expectation and the $0.95$-quantile for the Mat\'ern and the Gaussian kernel. The $0.95$-quantile does not guarantee that the estimate is included in the true excursion with probability $0.95$ in both examples. The conservative estimates instead are guaranteed to be inside the true excursion with probability $\alpha=0.95$. They correspond to Vorob'ev quantiles at levels $0.998$ (Mat\'ern) and $0.993$ (Gaussian). The conservative estimates were obtained with a $100 \times 100$ discretization of the unit square. Such high resolution grids lead to very high dimensional probability calculations. In fact, the dichotomy algorithm required $11$ computations of the probability $1-P_k(Q_{\rho^\prime} \subset \{\xi_x \leq t\})$ for each case. The discretization's size for $Q_\rho$ varied between $1213$ and $3201$ points in the Mat\'ern kernel case and between $1692$ and $2462$ points in the Gaussian case. Such high dimensional probabilities cannot be computed with the current implementation of the algorithm by Genz, however they could be computed with other Monte Carlo methods at higher computational costs. Instead, with the proposed method, the total computational time on a laptop with Intel Core i7 1.7GHz CPU and 8GB of RAM was equal to $365$ and $390$ seconds respectively for Mat\'ern and Gaussian kernel. In Supplementary Materials, Section~D, we compare the time required for conservative estimates when the core orthant probability estimate is computed with the GanMC, MET, GHK or QRSVN algorithms.

\section{Discussion}
\label{sec:Conclusion}
In this paper we introduced a new method to approximate high dimensional orthant Gaussian probabilities based on a decomposion of the probability in a low dimensional part $\pq$ and a remainder $\Rq$. 
%
The number of active dimensions $q$ and the dimensions themselves are chosen with two heuristic algorithms which provide good results in case of dense covariance matrix with anisotropic diagonal and anisotropic mean vector. An alternative proposal is choosing the first $q$ dimensions ordered according to the inverse Genz variable reordering proposed in \citet[Section~4.1.3]{Genz.Bretz2009}. While similar to the heuristics proposed here, this method is not efficient in high dimensions as it requires a full Cholesky decomposition of the covariance matrix. 
The remainder $\Rq$ is instead estimated with two methods: standard Monte Carlo and asymmetric nested Monte Carlo (anMC). Both methods showed higher efficiency than other state-of-the-art methods for dimensions higher than $1000$ when the orthant probability $\pi(t)$ is not a rare event in high dimensions. 

The version of the anMC method proposed here relies on the QRSVN algorithm to estimate $\pq$ and a rejection sampling to the estimate of $\Rq$. These choices, implemented as default in the package \verb|anMC|, can be easily changed to improve the method. In fact those particular choices do not prove to be the most efficient when $\pi(t)$ is very small in high dimensions, as shown in Appendix~\ref{sec:NumSmallProba}. For example, in supplementary material, we show that it is possible to implement the anMC method using the MET algorithm for the estimation of $\pq$ and a Hamiltonian Monte Carlo technique as truncated normal sampler. Such choices provide efficiencies close to the best state-of-the-art method in all cases and allow for efficient estimates in higher dimensions. 

Within its computational limits, the efficiency of $\pGANMC$ with default choices is mainly driven by the acceptance probability of the rejection sampler in $\RqANMC$, which in turn depends on $\pqW$. This highlights the existence of a trade-off between $\pqG$ and $\RqW$.  If the choice of $q$ and active dimensions is not optimal, then the acceptance probability of the rejection sampler becomes larger, making the estimation of $\Rq$ easier. An estimator $\pqW$ closer to $p$ makes the quantity $\Rq$ harder to estimate, however, in this case, $\RqANMC$ becomes more efficient than $\RqMC$ as the ratio between computational costs becomes more favourable.

In general, anMC relies on an initial step where several constants and probabilistic quantities are empirically estimated to choose the optimal $m^*$, the number of inner samples. In particular the cost parameters $\myc,\beta$, the slopes of the linear costs, might be hard to estimate if the constants $\myc_0, \alpha$ are comparatively large. In this case Algorithm~\ref{algo:Algo1} might not choose the optimal $m^*$. However, a numerical study of the algorithm behaviour for different choices of $m$ showed that, on the considered examples, even if the chosen $m$ is not optimal but it is close to optimal, the efficiency gain is very close to the optimal efficiency gain. 


The estimator $\pGANMC$ made the computation of conservative estimates possible for excursion sets of expensive to evaluate functions under general GRF priors. The $\RprogLang$~implementation of the algorithm is included in the package \href{http://www.github.com/dazzimonti/anMC}{\texttt{anMC}} currently available on CRAN and on GitHub.


\bigskip
\begin{center}
	{\large\bf SUPPLEMENTARY MATERIAL}
\end{center}

\begin{description}
	
	\item[SupplementaryMaterial: ] short description of the supplementary material provided, additional numerical results and computational times for Section~\ref{sec:ConservativeEsts}. (PDF file)
	
	\item[R-package anMC:] R-package implementing the GanMC, GMC procedures and the conservative estimates algorithm. (GNU zipped tar file)
	
	\item[R files (main):] The $\RprogLang$ files \verb|Section4_1.R|, \verb|Section4_2.R| and \verb|SuppMaterial_C.R| contain a blueprint of the code used for the numerical studies run in Section~\ref{sec:NumStudies}, Appendix~\ref{sec:NumSmallProba} and in Sections~C of the supplementary material. The code to generate the example in Section~\ref{sec:ConservativeEsts} is reproduced in \verb|Section5.R| and in \verb|SuppMaterial_D.R|. ($\RprogLang$ source code file)
	
	\item[R files (auxiliary):] The $\RprogLang$ file \texttt{create6dimGP.R} contains an auxiliary function to generate the examples used in Section~\ref{sec:NumStudies}. The $\RprogLang$ file \texttt{generateExamples.R} contains an auxiliary function to generate the realizations used in Section~\ref{sec:ConservativeEsts} and in Section~D, supplementary material. The $\RprogLang$ file \texttt{consEstGeneric.R} contains the function \verb|conservativeEstimate_generic| used for the computational comparison in Section~D, supplementary material. The $\RprogLang$ file \texttt{UserDefinedFunctions.R} defines the functions \verb|pmvnorm_usr| and \verb|trmvrnorm_usr| used in Section~C, supplementary material. ($\RprogLang$ source code file)
	
	\item[RData files:] The files \texttt{DataForGaussianFigure.RData} and \texttt{DataForMaternFigure.RData} contain the data to generate exactly Figure~\ref{fig:ConsEstimates}. (RData files)

\end{description}

\appendix

\section{Choice of active dimensions}
\label{sec:ActiveDims}

The estimator $\pqG$, introduced in Section~\ref{subsec:pq}, requires the choice of $q$, the number of active dimensions and the choice of the dimensions themselves. Algorithm~\ref{algo:AlgoSelq} describes the heuristic procedure implemented in \texttt{anMC} to select $q$ and obtain the active dimensions. Here we select $q$ by sequentially increasing the number of active dimensions until the relative change of $\pqG$ is less than the estimate's error.

\begin{algorithm}
	\SetKwInOut{Input}{Input}\SetKwInOut{Output}{Output}
	\Input{$q_0$, small initial $q$, e.g. $q_0=d^{1/3}$, and $q_\text{step}$ the increment of $q$, $\gamma> 0 $}
	\Output{$q$, $\pG[q]$}	
	Compute $\pG[q_0]$ and save $\operatorname{err}(\pG[q_0]) := 3\sqrt{\var(\pG[q_0])}$ \; 
	initialize $k=0$\;
	\Repeat{$\Delta(\pG[q_k]) < \gamma \operatorname{err}(\widehat{p}^G_{q_k})$ or $q_k>300$}{
		increment $k=k+1$ \;
		$q_k := q_0+k q_\text{step}$ \;
		choose $q_k$ active dimensions, compute $\widehat{p}^G_{q_k}$ and $\operatorname{err}(\pG[q_k])$ \;
		compute $\Delta(\pG[q_k]) = \frac{\abs{\pG[q_k] - \pG[q_{k-1}]}}{1+\pG[q_k]}$ \;
	}
	$q=q_k$ and $\pqG=\pG[q_k]$\;
	
	\caption{Select $q$, active dimensions and compute $\pqG$.}
	\label{algo:AlgoSelq}
\end{algorithm}

The constant $\gamma >0$ is chosen equal to $1$ in our implementation. Moreover the algorithm stops if $q_k>300$ to avoid using Genz's algorithm in high dimensions.

\subsection{Add spatial information}
If the random vector $X$ comes from a GRF discretized over a set of points $E_{\text{spat}} =\{e_1, \dots, e_d\} \subset \R^l$, then we can exploit this information to choose $E_q$.
Let us consider the sequence of vectors $(\boldsymbol{\delta}_j)_{j=1, \dots, q}$, defined for each $j$ as
\begin{equation*}
\boldsymbol{\delta}_j =
\prod_{k=1}^{j}\operatorname{dist}(e_{i_k},E_{\text{spat}}) \quad (j=1,\ldots, q)
\end{equation*}
where $\operatorname{dist}(e_{i_k},E_{\text{spat}})$ denotes the $d$-dimensional vector of Euclidean distances between $e_{i_k}$ and each point in $E_{\text{spat}}$ and $\{e_{i_1}, \dots, e_{i_q}\}$ are the points corresponding to the selected active dimensions $E_q$. We then adjust Methods A, B by sampling the $j$th active dimension with probabilities given by the component-wise products $p_t\frac{\boldsymbol{\delta}_j}{\lVert \boldsymbol{\delta}_j \rVert }$ and $p_t(1-p_t)\frac{\boldsymbol{\delta}_j}{\lVert \boldsymbol{\delta}_j \rVert }$ respectively.

\section{Proofs}
\label{sec:proofs}

\textbf{Proof of Proposition~\ref{pro:VarHatp}}
\begin{proof}
	We have that $\E[\pqW] = \pq$ and $\E[\RqW] = \Rq$. Then we have
	\begin{equation}
	\var(\widehat{p}) = \var(\pqW) +  \underbrace{\var((1-\pqW)\RqW)}_{=\blacksquare} + 2\underbrace{\cov(\pqW,(1-\pqW)\RqW)}_{=\blacktriangle}.
	\label{eq:proofVar1st}
	\end{equation}
	We can write the variance~$\blacksquare$ and the covariance~$\blacktriangle$ as
	\begin{align*}
	\blacksquare=\var((1-\pqW)\RqW) &= (1-\pq)^2\var(\RqW)+\Rq^2\var(\pqW)+\var(\pqW)\var(\RqW), \\
	\blacktriangle=\cov[\pqW,(1-\pqW)\RqW] &= -\var(\pqW)\Rq,
	\end{align*}
	respectively, by exploiting the independence of $\pqW$ and $\RqW$. By plugging in those expressions in Equation~\eqref{eq:proofVar1st} we obtain the result in Equation~\eqref{eq:VarHatp}.
\end{proof}

\textbf{Proof of Proposition~\ref{pro:general}}
\begin{proof}
	\begin{align} \nonumber
	\var(\widetilde{G}) &= \frac{1}{n^2m^2}\var\left(\sum_{i=1}^n\sum_{j=1}^m g(W_i,Z_{i,j})\right)  
	= \frac{1}{nm^2}\var\left(\sum_{j=1}^m g(W_1,Z_{1,j})\right) \\ \nonumber
	&= \frac{1}{nm^2}\sum_{j=1}^m \sum_{j^\prime=1}^m \cov\big(g(W_1,Z_{1,j}),g(W_1,Z_{1,j^\prime})\big) \\ \nonumber
	&= \frac{1}{nm^2}\bigg[m \var(g(W_1,Z_{1,1})) + m(m-1) \cov(g(W_1,Z_{1,1}),g(W_1,Z_{1,2})) \bigg] \\
	&= \frac{1}{nm^2}\left[ m\var(g(W_1,Z_{1,1})) + m(m-1)\blacklozenge  \right].
	\label{eq:varDecom1}
	\end{align}
	where the first equality is a consequence of the independence of $W_1, \dots, W_n$ and the third equality is a consequence of the independence of $Z_{i,j}$ and $Z_{i,j^\prime}$ conditionally on $W_i$. Moreover the covariance denoted by $\blacklozenge$ in~\eqref{eq:varDecom1} can be written as follows.
	\begin{align}
	\nonumber \blacklozenge &= \underbrace{\E\big[\cov (g(W_1,Z_{1,1}),g(W_1,Z_{1,2}) \mid W_1) \big] }_{=0 \ Z_{1,1}, Z_{1,2}\text{ independent conditionally on } W_1 } + \underbrace{\cov\big( \E[g(W_1,Z_{1,1}) \mid W_1], \E[g(W_1,Z_{1,2}) \mid W_1] \big)}_{= \var\left(\E[g(W_1,Z_{1,1}) \mid W_1]\right)}  \\
	&= \var\big(\E[g(W_1,Z_{1,1}) \mid W_1]\big) = \var\big( g(W_1,Z_{1,1}) \big) - \E\bigg[ \var\big( g(W_1,Z_{1,1}) \mid W_1 \big) \bigg].
	\label{eq:DecomLozange}
	\end{align}
	Equations~\eqref{eq:varDecom1} and~\eqref{eq:DecomLozange} give the result~\eqref{eq:varianceDecomp}.
	
\end{proof}

\textbf{Proof of Corollary~\ref{cor:OptimNum}}
\begin{proof}
	Denote with $e=\beta(A-B),\ f=(\alpha +\myc)(A-B) + \beta B,\ g= (\myc+\alpha)B,\ h=C_\text{tot}-\myc_0$, then 
	\begin{equation}
	\var(\widetilde{G})(m) = \frac{em^2 + fm+ g}{hm}.
	\label{eq:appendixVarG}
	\end{equation}
	Observe that the first and second derivatives of $\var(\widetilde{G})$ with respect to $m$ are respectively
	\begin{equation*}
	\frac{\partial \var(\widetilde{G})}{\partial m} = \frac{1}{h} \left[ e - \frac{g }{m^2} \right],   \quad
	\frac{\partial^2 \var(\widetilde{G})}{\partial m^2} = \frac{2g }{h m^3}. 
	\end{equation*}
	The second derivative is positive for all $m>0$ then $\var(\widetilde{G})$ is a convex function for $m>0$ and the point of minimum is equal to the zero of $\partial\var(\widetilde{G})/\partial m$, which is $m = \sqrt{g/e}= \widetilde{m}$.
	
	Since $\var(\widetilde{G})$ is convex in $m$, the integer that realizes the minimal variance is either $\lfloor \widetilde{m} \rfloor$ or $\lceil \widetilde{m} \rceil$. By plugging in $m=\widetilde{m}-\varepsilon=\sqrt{g/e} -\varepsilon$ and $m=\widetilde{m}-\varepsilon +1 =\sqrt{g/e} -\varepsilon+1$ in Equation~\eqref{eq:appendixVarG}, we obtain the condition in~\eqref{eq:approxMstar}.
\end{proof}

\textbf{Proof of Proposition~\ref{pro:comparisonVar}}
\begin{proof}
	First of all notice that the total cost of sampling $\widehat{G}$ is $C_{\text{tot}} = \myc_0 + n(\myc + C_{Z \mid W}) = \myc_0 n(\myc + \alpha + \beta)$. By isolating $n$ in the previous equation we obtain $n= \frac{\overline{C}_{\text{tot}}}{\myc + \alpha + \beta}$, where $\overline{C}_{\text{tot}}$ for the sake of brevity and, by computations similar to those in Proposition~\ref{pro:general} we obtain
	\begin{equation*}
	\var(\widehat{G})= \frac{\myc + \alpha + \beta}{\overline{C}_{\text{tot}}}\var(g(W_1,Z_{1,1})) = \frac{\myc + \alpha + \beta}{\overline{C}_{\text{tot}}} A,
	\end{equation*}
	where $A=\var(g(W_1,Z_{1,1}))$. In the following we will also denote $B=\E\big[ \var( g(W_1,Z_{1,1}) \mid W_1 ) \big]$ as in Corollary~\ref{cor:OptimNum}.
	Let us now substitute $N_{C_{\text{fix}}}(m^*)$ in equation~\eqref{eq:varianceDecomp}, thus obtaining
	\begin{align} \nonumber
	\var(\widetilde{G}) &= \frac{(\myc +\alpha + \beta m^*)Am^* - (m^*-1)(\myc +\alpha +\beta m^*)B}{\overline{C}_\text{tot}m^*} \\ \nonumber
	&=\var(\widehat{G})\frac{(m^*)^2\beta(A-B) + m^*[(\myc +\alpha)(A-B) +\beta B] +(\myc +\alpha)B}{A(\myc +\alpha +\beta)m^*} \\
	&= \var(\widehat{G})\frac{2(\alpha +\myc)B + m^*[(\myc +\alpha)(A-B) +\beta B] }{A(\myc +\alpha +\beta)m^*},
	\label{eq:varOptim1}
	\end{align}
	where in~\eqref{eq:varOptim1} we substituted $(m^*)^2$ from Corollary~\ref{cor:OptimNum}. By rearranging the terms, we obtain
	\begin{equation*}
	\var(\widetilde{G}) 
	= \var(\widehat{G})\left[ 1- \frac{(m^*-2)(\myc +\alpha)B +m^*\beta (B-A)}{A(\myc +\alpha +\beta)m^*} \right] = \var(\widehat{G})\left[ 1- \eta \right].
	\end{equation*}
	Since $A-B, B, \myc, \beta,\alpha$ are always positive, then $\eta<1$ for all $m^*>0$. Moreover $\eta>0$ if
	\begin{equation*} 
	m^* > \frac{2(\alpha+\myc)B}{(\alpha+\myc)B + \beta(A-B)}.
	\end{equation*}
\end{proof}

\section{Numerical study for small and large probabilities}
\label{sec:NumSmallProba}

The algorithm GanMC was originally developed to estimate probabilities in the form $\pi(t)$ for high dimensional problems where this probability does not decrease rapidly with the dimension, as, for example, in the numerical study shown in Section~\ref{sec:NumStudies}. In this section we study two limit cases where this assumption is challenged: first we consider very high probability values for $\pi(t)$, second we consider the very small probability case. For very small probabilities, the anMC method with the default choices implemented in \verb|anMC| might not be the correct choice. In fact, if $\pi(t)$ becomes too small then $p=1-\pi(t)$ is very close to $1$ and $\pq$ is very close to $p$. As suggested in Section~\ref{subsec:MCforRq}, the residual term $\Rq$ becomes increasingly hard to estimate with rejection sampling as $\pq$ becomes small. An alternative sampler for truncated normal vectors might improve performance. This can be achieved in the function \verb|ProbaMax| of the package \verb|anMC| by choosing a user defined truncated normal sampler. 

\begin{figure}
	\begin{subfigure}{0.495\textwidth}
		\centering
		\includegraphics[width=\linewidth]{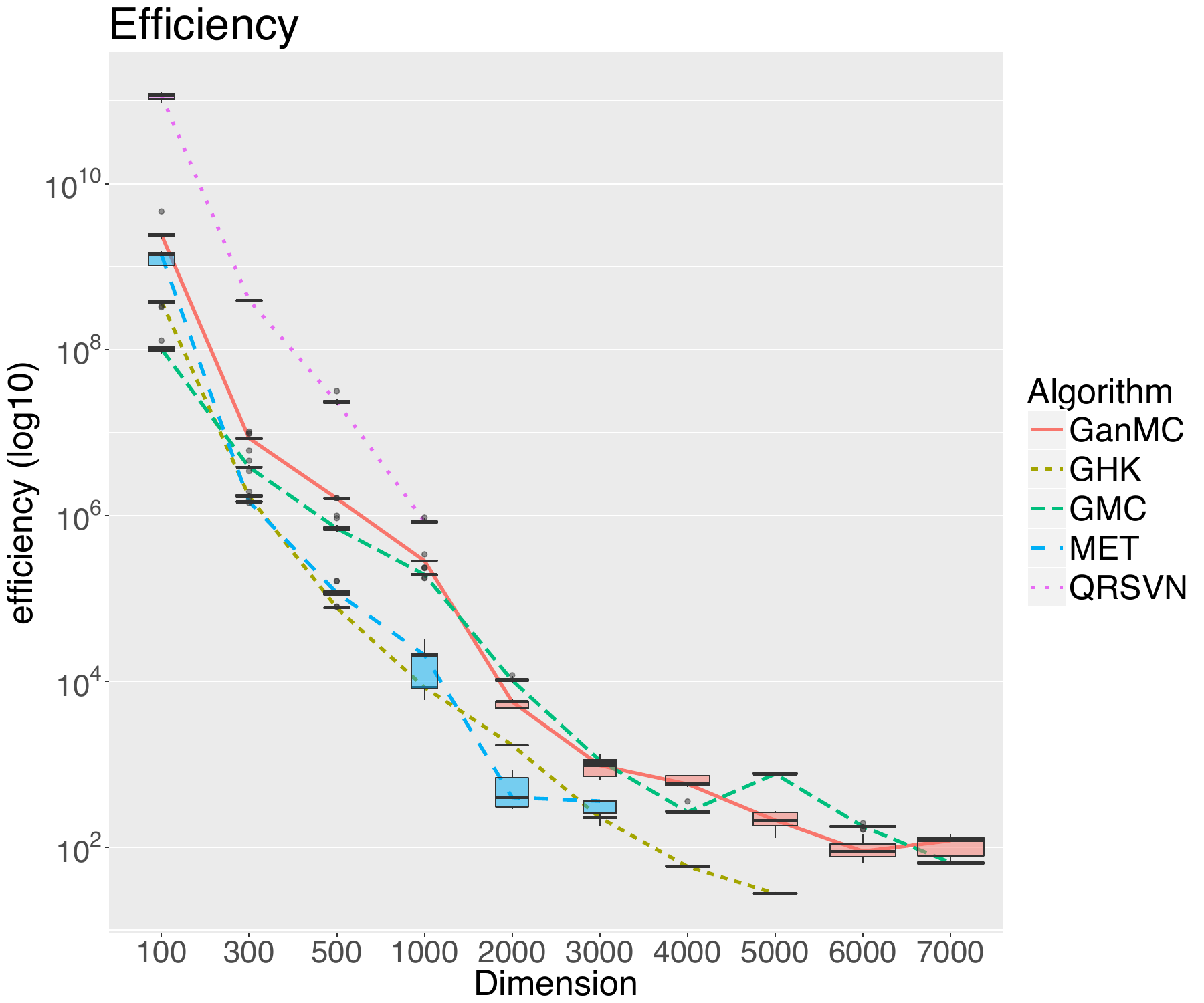}
		\caption{High $\pi(t)$ state, $t=7.5$. The median estimated value for $p=1-\pi(t)$ ranges from $0.00315$ to $0.32564$.}
		\label{fig:CompEff6DThresh7_5}
	\end{subfigure} \hspace{0.01\textwidth}
	\begin{subfigure}{0.495\textwidth}
		\centering
		\includegraphics[width=\linewidth]{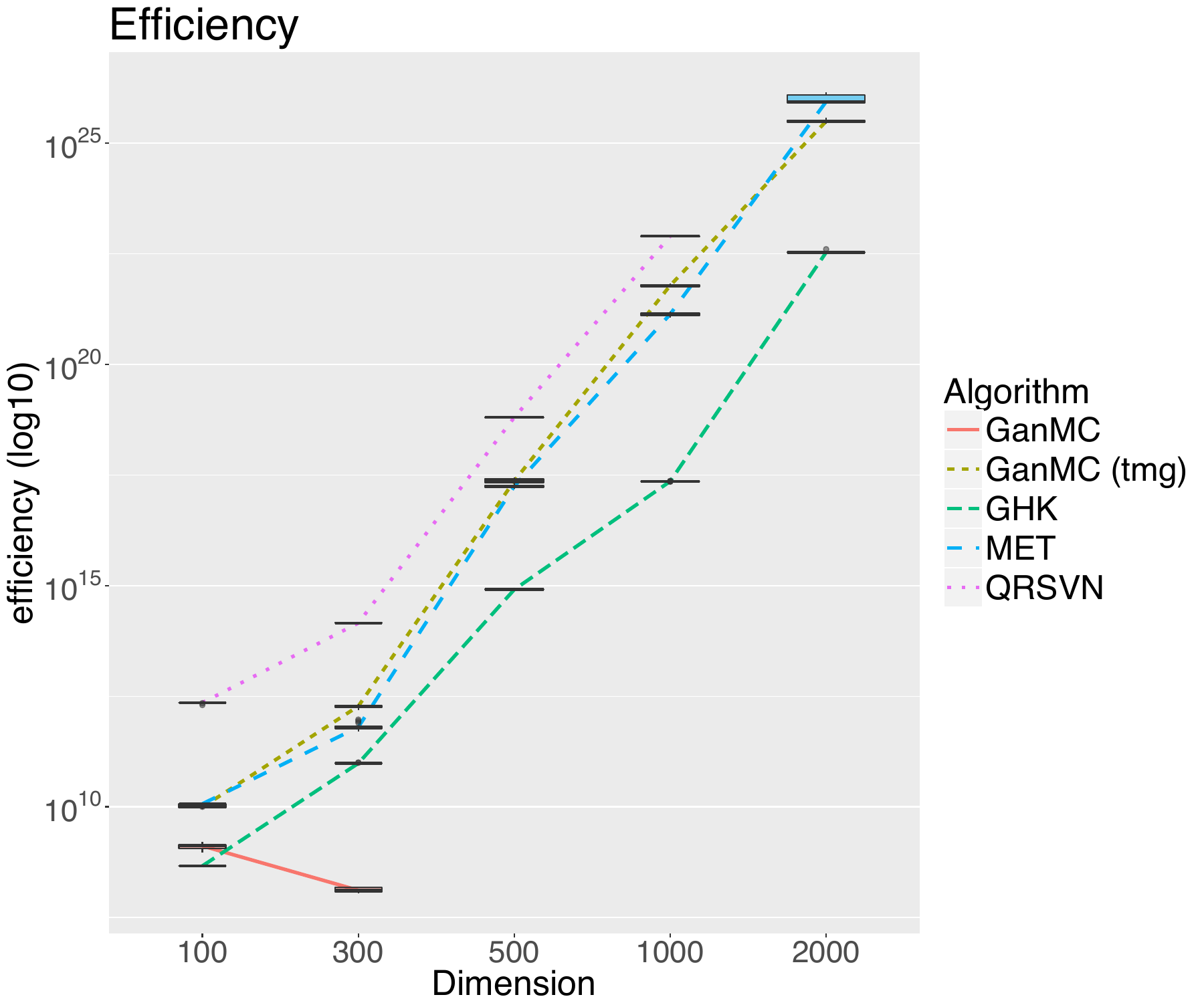}
		\caption{Efficiency for the computation of $\pi(t)$ with $t=3$. The median estimated value for $\pi(t)$ ranges from $9.02\times 10^{-3}$ to $1.33\times10^{-13}$.}
		\label{fig:CompEff6DThresh3}
	\end{subfigure} 
	\caption{Efficiency of the probability estimator versus the dimension $d$. For each $d$ the experiment is reproduced $15$ times. Values in logarithmic scale.}
	\label{fig:compEff6D_3}
\end{figure}

We construct two benchmark studies with the problem defined in Section~\ref{subsec:NumInner} and by changing the threshold $t$ to $t=7.5$ for the high probability values and $t=3$ for the small values. 

In Figure~\ref{fig:compEff6D_3} we show an efficiency comparison for estimating $\pi(t)$, with $t=3$ and $t=7.5$, with the algorithms GanMC, GMC, GHK, MET and QRSVN, see Section~\ref{subsec:CompSoA} for details.

The threshold $t=7.5$ leads to a high probability setup as the median estimate for $\pi(t)$ ranges between $0.99685$, for $d=100$, and $0.67436$, for $d=7000$. Figure~\ref{fig:CompEff6DThresh7_5} compares the estimators' efficiency. In this case the QRSVN is the most efficient algorithm in low dimensions. The GMC and the GanMC algorithms however are the most efficient for all dimensions higher than $2000$. The GanMC algorithm is $3$ times more efficient than the MET for $d=3000$ and $9$ times more efficient than GHK for $d=5000$. In this setup the computational cost of the rejection sampler in $\RqANMC$ is not much higher than the conditional sampler. In fact, the acceptance probability of the rejection sampler is always higher than $0.6$. In most replications this leads to a choice of $m^*$ very small or even equal to $1$. Thus GanMC is slower than GMC because of Part 1 in Algorithm~\ref{algo:Algo1} while achieving the same variance. This is the main reason why the GMC algorithm proves to be more efficient in dimensions $2000, 3000, 5000, 6000$, in fact for $d=5000$ GMC is $3.6$ times more efficient than GanMC and for $d=6000$ the ratio is $1.9$. Computations for this experiment were carried on the cluster of the University of Bern on machines with Intel Xeon CPU 2.40GHz and 16 GB RAM. 

For both the case $t=7.5$ and $t=5$ the probability $\pi(t)$ does not qualify as small. Figure~\ref{fig:CompEff6DThresh3} shows an efficiency comparison for $t=3$, where the value of $\pi(t)$ is equal to $9.02\times 10^{-3}$ for $d=100$, decreases to $1.33\times 10^{-13}$ for $d=2000$ and becomes too small to be estimated reliably for larger dimensions. The GHK method while performing well in low dimensions is not able to estimate reliably the probability for $d=1000$ or higher. Notice that the GanMC method with the default rejection sampler for the residual part does not work when $d>500$ because the acceptance probability is too low. If the rejection sampler is replaced by the Hamiltonian Monte Carlo method for truncated normal vectors described in \citet{PakmanPaninski2014} and implemented in the $\RprogLang$~package \verb|tmg|, then the GanMC method performs better than GHK and MET for dimensions lower that $2000$, however it is still $3$ times less efficient than MET for $d=2000$. Higher dimensional comparison were not possible as MET and GHK method required more memory than allowed. Computations for the experiment $t=3$ were carried out on the Idiap Research Institute computing grid, on machines with Intel Xeon E312xx (Sandy Bridge) CPU 3.00GHz and 8 GB RAM.

\bibliographystyle{apalike}
\bibliography{biblioArticle}

\end{document}